\documentclass[footinbib,twocolumn,showpacs,amsmath,amstex,amssymb,mathfonts,superscriptaddress,prx]{revtex4}
\usepackage{graphicx}
\usepackage{color}
\usepackage{bm}

\usepackage{graphicx}
\usepackage{color}
\usepackage{bm}
\usepackage{amsmath}
\usepackage{amssymb}
\usepackage{amsthm}
\usepackage{amsfonts}
\usepackage{multirow}
\usepackage{makecell}
\usepackage{float}
\usepackage{comment}
\usepackage{enumitem}
\usepackage{verbatim}
\usepackage{mathtools}

\usepackage{bbm}

\begin{document}

\title{Quantum geometric fluctuations in fractional quantum Hall fluids}
\author{Bo Yang} 
\affiliation{Division of Physics and Applied Physics, Nanyang Technological University, Singapore 637371.}
\pacs{73.43.Lp, 71.10.Pm}

\date{\today}
\begin{abstract}
We present here a complete microscopic theory of a family of neutral excitations in the fractional quantum Hall fluids, related to the geometric fluctuations of the quantum Hall ground states. Many of the physical properties of such geometric modes can be inferred analytically. These include the chirality, multiplicity and energy of these geometric modes, as well as the relationship to the density modulation of the ground states of both incompressible and compressible fluids, with or without translational symmetry (e.g. the bubble and stripe phases). With a particular focus on the recently experimentally measured graviton modes as the special case, we elucidate both the universal aspects of the geometric modes and the non-universal aspects dependent on the details of the microscopic Hamiltonians. The microscopic theory explains some of the phenomenological components in the effective field theory and composite fermion theory. It predicts how geometric or graviton modes of both chiralities can be measured in experiments for any topological or compressible phases at different energy scales. In particular we show gapped geometric modes can exist even for compressible FQH phases, though translational symmetry of the ground state is important for such modes to couple to external probes (e.g. Raman scattering) in the long wavelength limit.
\end{abstract}

\maketitle 

\section{Introduction}\label{sec_intro}
Strongly correlated quantum fluid confined within a two-dimensional manifold with broken time-reversal symmetry is one of the most interesting condensed matter systems with exotic collective excitations emerging from unconventional phases of matters. The prototypical examples are the fractional quantum Hall (FQH) effect\cite{prange} and its lattice analog of the fractional quantum anomalous Hall effect (i.e. the fractional Chern insulators)\cite{regnault2011fractional,chang2013experimental,lu2024fractional}, with a two-dimensional electron gas at low temperature with intrinsic non-vanishing average Berry curvature in a flat band. The charged collective excitations in gapped topological quantum fluids have been extensively studied due to their anyonic or even non-Abelian statistical properties. These topologically protected quasiparticles have no classical analog, each carrying a localised fraction of a single elementary electrical charge, paradoxically via the quantum coherence over a large number of electrons\cite{wilczek1982quantum,halperin1984statistics,arovas1984fractional,leinaas2002spin, stern2008anyons,feldman2021fractional}. The incompressibility gap, determined by the energy gap of charged excitations with electron density greater than that of the ground state, needs to be larger than the temperature for the universal topological properties to be robust, a necessary condition for these charged quasiparticles to store and manipulate quantum information that can be very resistant against quantum decoherence\cite{preskill1998fault,sarma2005topologically,nayak2008non}.

Neutral excitations in these strongly correlated quantum fluids started with the seminal papers by Girvin, Macdonald and Platzmann (GMP) nearly three decades ago\cite{girvin1985collective, girvin1986magneto}, and were also extensively studied in the framework of the composite fermion (CF) theory and with inelastic light scattering\cite{majumder2009collective, roncaglia2010pfaffian,pinczuk1993observation}. For the incompressible topological phases, the simplest neutral excitation consists of a pair of well separated quasiparticles of the opposite charge (i.e. the quasihole-quasielectron pair). It carries a non-zero dipole moment and a finite energy gap equivalent to the sum of the quasihole and quasielectron energy, which is still the charge gap. A fundamentally different type of neutral excitations have vanishing dipole moment and can no longer be viewed as a composite ``exciton-like" quasihole-quasielectron pair\cite{PhysRevLett.108.256807}. The energies of such neutral excitations are also independent energy scales not necessarily related to the charge gap. It has been proposed more than a decade ago that one family of such neutral excitations in the long wavelength limit can be understood as a two-dimensional version of non-relativistic gravitons, from the quantization of the ground state metric fluctuations\cite{Haldane2011,yang2012band, son2013newton, Yang2013, liang2024evidence, Golkar2016, Luo2016, yang2016acoustic, Gromov2017, wang2021analytic, wang2023geometric}. Recent studies also show such neutral excitations can act as force-mediating particles capturing both the transition from the fully gapped FQH phases to the nematic FQH phases, and the exotic dynamics of the fractionalization of anyonic excitations in the neighbourhood of such transitions\cite{PhysRevLett.127.046402}. 

Gravitons and related neutral excitations emerging from the topological quantum fluids are the main focuses of this work, motivated by the recent success in experiments in the detection of chiral gravitons in a number of FQH states with inelastic Raman scattering\cite{liang2024evidence}. This important progress leads to an entirely new probe for the interplay of the geometric and topological properties in the FQH fluids, and for the measurement of this family of excitations that potentially play many roles in the dynamics of strongly correlated fluids. It is however still a bit controversial about why we ``invent" the terminology of ``gravitons" for this family of neutral excitations, and such excitations are not easy to understand from some of the common approaches in FQH physics. The effective fieldtheory\cite{Golkar2016,golkar2016higher,maciejko2013field,PhysRevLett.119.146602,PhysRevX.7.041032,PhysRevB.97.195103,Nguyen2021} approach couples the quantum Hall fluids to the metric deformation of the Hall manifold in a phenomenological manner, but without detailed microscopic input it cannot predict when the microscopic sum rules (e.g. the Haldane bound) would be violated and how many additional ``graviton modes" need to be added by hand. The composite fermion (CF) and the related parton theory\cite{jain1989composite, PhysRevB.40.8079,jain2015composite} have been very successful in the construction of trial wavefunctions for the many-body FQH states with transparent physical interpretations. It is however interesting to note that there is no geometric interpretation of the quantum Hall fluids within the CF or parton theory themselves; all neutral excitations of the FQH fluids at small and large wave vectors are described as the CF or parton excitons in a unified manner\cite{golkar2016higher, balram2021highenergy, nguyen2021multiple, nguyen2022multiple,nguyen2023supergravity,balram2024splitting}. While the exciton descriptions agree with the exact numerical diagonalization well in most cases for Coulomb like interactions, the additional physics relating to the emergent quantum geometry for neutral excitations at \emph{small wave-vectors} is not explicitly captured. It thus could be both conceptually and technically challenging to understand the fundamental aspects of gravitons (and other geometric modes), as well as their potential connections to 2D quantum gravity and string theory (e.g. via the $W_\infty$ algebra), from the perspective of CF/parton excitons.

The work aims to give a modern perspective of the geometric modes of the FQH fluids, with a particular focus on the graviton modes as the special cases. In particular, we hope to show that the rigorous microscopic formulation of such neutral excitations can also be simplest to understand their dynamical and geometric properties, as well as the usefulness of their experimental measurements for extracting important information about the underlying phases of matter. The organization of the paper is as follows: In Sec.~\ref{sec_density} we give a short review of the density modes in FQH fluids, and clarify the relations between different terminologies such as the single mode approximation and the GMP modes; in Sec.~\ref{sec_lwl} we review the derivation of the quantum metric aspects of the long wavelength density modes and the associated Haldane bound\cite{haldane2009hall}, and for the first time introduce a complete family of geometric modes describing the general shape fluctuation of a generic many-body state; in Sec.~\ref{sec_chiral} we give the microscopic derivation of the origin of the graviton chirality (which can be readily generalized to all geometric modes), showing that gravitons of both chiralities exist for a general quantum fluids (e.g. the edge gravitons on the disk geometry) at different energy scales; In Sec.~\ref{sec_dynamics} we go beyond our recent work\cite{wang2023geometric} to show a rigorous formulation of the so-called ``multiple gravitons" that clarifies some of the misunderstandings in the literature, with the associated construction of model Hamiltonians showing this multiplicity is dynamical and dependent on the details of the interaction; in Sec.~\ref{sec_chs} we go beyond the realistic interaction discussed in Sec.~\ref{sec_dynamics}, and give a more general formulation of the graviton multiplicity within the framework of the conformal Hilbert spaces\cite{yang2021gaffnian,yang2022anyons, yang2022composite, trung2023spin, wang2021analytic, wang2023geometric}; in Sec.~\ref{sec_exp} we discuss the implications and predictions of the experimental measurements of geometric modes from our microscopic theory, and in Sec.~\ref{sec_connection} we first briefly discuss about the phenomenological theories of the gravitons from the perspective of the more rigorously formulated microscopic theory; this is followed by a detailed discussion about the other microscopic formulation of the gravitons with two-body and multi-body operators.

\section{The density modes of the FQH fluids}\label{sec_density}

We start with a brief historical review and to set up the notations for the rest of the paper. The seminal work by Girvin, Macdonald and Platzmann (GMP) aimed to explain an interesting branch of neutral excitations of the Laughlin state at filling factor $\nu=1/3$, as evidenced from the numerical computations\cite{girvin1985collective, girvin1986magneto}. Inspired by Feynman's treatment of the collective excitations in superfluid $^4$He\cite{feynman2018statistical}, it was proposed that a family of trial wavefunctions can be constructed with the electron density operator projected into a single Landau level (LL). All physics of the graviton modes, as well as the higher spin modes we will discuss later, is fundamentally (and in many ways in a very simple manner) from the projected electron density operator. The density operator is also the most relevant in experiments when the external probe couples to the modulation of the electron density. It is thus worthwhile to give it a more detailed treatment. 

\textit{The projected density operator--}The bare density operator in real space is given by $\hat\rho(\bm r)=\sum_i\delta^2\left(\bm r-\hat{\bm r_i}\right)$, where $\bm r=(r^x,r^y)$ is the real space coordinate vector and $i$ is the electron index with $\hat{\bm r}_i=\left(\hat r^x_i,\hat r^y_i\right)$. In the presence of the magnetic field $\bm B_z$, the magnetic length is given by $\ell_B=\sqrt{\hbar/e\bm B_z}$, where $e$ is the elementary electron charge. We can thus always express $\hat r^a_i=\tilde R^a_i+\bar R^a_i$, where $\tilde R_i^a$ is the cyclotron coordinate operator and $\bar R_i^a$ is the guiding center coordinate operator. The LLs are flat bands giving the energy levels of the kinetic energy Hamiltonian $\hat H_0\left(\tilde R_i^a\right)$; it in general can be quite complicated in realistic materials (so the energy levels are not necessarily equally spaced) but only depends on $\tilde R_i^a$ and not on $\hat r_i^a$ or $\bar R_i^a$. It is important to note the following commutation relations:
\begin{eqnarray}
&&[\tilde R_i^a,\tilde R_j^b]=i\ell_B^2\delta_{ij}\epsilon^{ab}, [\bar R_i^a,\bar R_j^b]=-i\ell_B^2\delta_{ij}\epsilon^{ab}\label{com1}\\
&&\qquad\qquad\qquad [\tilde R_i^a,\bar R_j^b]=0\label{com2}
\end{eqnarray}
where $\delta_{ij}$ is the Kronecker delta and $\epsilon^{ab}$ is the antisymmetric tensor. In particular, Eq.(\ref{com2}) implies $[\hat H_0,\bar R_i^a]=0$, a relationship directly responsible for the large degeneracy within each kinetic energy and thus the LLs as flat bands. The commutation relations above establishes that $\tilde R_i^a$ physically moves electrons between different LLs, while $\bar R_i^a$ moves electrons within a single LL. Note that all these algebraic structures are established without the need to pick a specific gauge for the external electromagnetic field. 

In the momentum space, the bare density operator is given by $\hat\rho_{\bm q}=\sum_ie^{iq_a\hat r_i^a}$ with the momentum vector $\bm q=(q_x,q_y)$. In this work, we focus on the physics entirely within a single LL, so $\tilde R_i^a$ will be completely ignored unless otherwise specified. It is proposed by GMP that the guiding center density operator projected into a single LL, $\bar\rho_{\bm q}=\sum_ie^{iq_a\bar R_i^a}$, which is the physically relevant operator for electron density within the Hilbert space of a single LL, can be used to generate density modes as neutral excitations in the FQH systems\cite{girvin1985collective, girvin1986magneto}. Indeed for systems with translational and rotational invariance (e.g. clean systems in the thermodynamic limit), any neutral excitation with quantum number $\bm q$ can be expressed as a linear combination of the family of states given below:
\begin{eqnarray}
|\psi_{\bm q}^{\bm q_1,\bm q_2\cdots\bm q_n}\rangle\sim \bar\rho_{\bm q_1}\bar\rho_{\bm q_2}\cdots\bar\rho_{\bm q_n}|\psi_0\rangle\label{dmode}
\end{eqnarray}
with $\bm q=\sum_{i=1}^n\bm q_i$ and $|\psi_0\rangle$ is the ground state. From Eq.(\ref{com1}) it is also easy to show the famous GMP algebra of the guiding center density operator as follows:
\begin{eqnarray}
[\bar\rho_{\bm q_1},\bar\rho_{\bm q_2}]=2i\sin\left(\frac{\epsilon^{ab}q_{1a}q_{2b}}{2}\ell_B^2\right)\bar\rho_{\bm q_1+\bm q_2}\label{gmp}
\end{eqnarray}
a closed algebra capturing the defining symmetry of the quantum Hall systems. The GMP algebra is isomorphic to the $W_\infty$ algebra, connecting to the physics of conformal field theory and string theory\cite{bergshoeff1990area,  sezgin1992area, cappelli1993infinite,flohr1994infinite, parameswaran2012fractional,cappelli2021w, wang2024closed}. Such connections very much manifest themselves via the emergence of neutral excitations as we will discuss in the later parts.

\textit{The single mode approximation--}Instead of looking at multiple density mode excitations given in Eq.(\ref{dmode}), GMP proposed to focus on a family of trial wavefunctions which are a small subset of states in Eq.(\ref{dmode})
\begin{eqnarray}\label{sma}
|\psi^{\text{SMA}}_{\bm q}\rangle\sim\bar\rho_{\bm q}|\psi_0\rangle
\end{eqnarray}
consisting of an excitation of a single density mode from the ground state. This so-called single mode approximation (SMA) successfully captures the essential physics of the branch of neutral excitations for small $\bm q$, and in some cases (e.g. the Laughlin\cite{laughlin1983anomalous} and Moore Read states\cite{moore1991nonabelions}) all the way up to $|\bm q|\sim\ell_B^{-1}$. In analogy to Feynman's description of the roton modes in $^4$He, the branch of neutral excitations found in the numerics is named the magnetoroton modes. In contrast the SMA given in Eq.(\ref{sma}) is called the GMP mode\cite{girvin1986magneto}.

It is useful to clarify the terminology here to avoid confusions. The magnetoroton modes are the physical neutral excitations in FQH fluids. The term of SMA modes and GMP modes are interchangeable, describing the same family of trial wavefunctions approximating the magnetoroton modes, and they are only good approximations for $|\bm q|\lesssim\ell_B^{-1}$. Moreoever one can prove in the long wavelength limit $|\bm q|\to 0$, the GMP or SMA modes are \emph{identical} to the magnetoroton modes\cite{PhysRevLett.108.256807}. Thus only in the limit of $|\bm q|\to 0$, all three terms of SMA, GMP and magnetoroton modes are interchangeable. It is also in this limit the physics of the neutral excitations are particularly interesting, as they can be experimentally measured, and analytically described by a quantum geometric theory.

\section{The long wavelength limit}\label{sec_lwl}

The GMP modes capture the physics of the magnetoroton modes in the long wavelength limit, but it was not noticed back then the physics in this limit is geometric. Haldane first pointed out there is a geometric degree of freedom of the incompressible FQH ground state, parametrized by a unimodular metric that can be deformed and rotated\cite{haldane2009hall, Haldane2011}. The interesting connection of this metric deformation to the long wavelength GMP mode can be revealed by focusing on the following trial state:
\begin{eqnarray}\label{graviton}
|\psi_{|\bm q|\to 0}\rangle=\lim_{\bm q\to 0}\frac{1}{\sqrt S_{\bm q}}\delta\bar\rho_{\bm q}|\psi_0\rangle
\end{eqnarray}
where the subtlety is we use the regularized guiding center density operator $\delta\bar\rho_{\bm q}=\bar\rho_{\bm q}-\langle\psi_0|\bar\rho_{\bm q}|\psi_0\rangle$. We obviously have $\delta\bar\rho_{\bm q}=\bar\rho_{\bm q}$ for $|\bm q|>0$ so it only affect the case of $|\bm q|\to 0$, with the purpose of guaranteeing $\langle\psi_0|\psi_{|\bm q|\to 0}\rangle=0$. Note that the state in Eq.({\ref{graviton}) is normalised by the guiding center structure factor $S_{\bm q}=\langle\psi_0|\delta\bar\rho_{-\bm q}\delta\bar\rho_{\bm q}|\psi_0\rangle$ which vanishes as $|\bm q|^4$ in the limit of $|\bm q|\to 0$, so the state is well defined at $\bm q=0$ even though the unnormalized $|\psi'_{|\bm q|\to 0}\rangle=\lim_{\bm q\to 0}\delta\bar\rho_{\bm q}|\psi_0\rangle$ vanishes.

To see the geometric connection, it is useful to look at the unnormalized state $|\psi'_{|\bm q|\to 0}\rangle$ and expand $\bm q$ order by order in the long wavelength limit, with the following definition:
\begin{eqnarray}\label{expand}
&&|\psi'^{\left(n\right)}\rangle=\sum_{k=1}^n\frac{i^k}{k!}q_{a_1}q_{a_2}\cdots q_{a_k}\bar \Lambda^{a_1a_2\cdots a_k}|\psi_0\rangle\\
&&\bar\Lambda^{a_1\cdots a_k}=\sum_i\bar R_i^{a_1}\cdots \bar R_i^{a_n}-\langle\psi_0|\bar R_i^{a_1}\cdots \bar R_i^{a_n}|\psi_0\rangle\label{generator}\quad\quad
\end{eqnarray}
In Eq.(\ref{generator}) we can also symmetrize the upper indices (which we will assume to be the case onwards) and it guarantees $\langle\psi_0|\psi'^{\left(n\right)}\rangle=0$ order by order.

If the ground state $|\psi_0\rangle$ is translationally and rotationally invariant, then $|\psi'^{(1)}\rangle$ vanishes because $\hat P=\sum_i\bar R_i^a$ is the center of mass magnetic translation generator. This is the case for incompressible FQH fluids or the composite Fermi liquid (CFL), and it is a point we will come back later. Thus for these quantum fluids we have the following:
\begin{eqnarray}
|\psi_{|\bm q|\to 0}\rangle\sim|\psi'^{(2)}\rangle=-\frac{1}{2}q_aq_b\bar\Lambda^{ab}|\psi_0\rangle
\end{eqnarray}
It was pointed out by Haldane that the rank-2 tensor $\bar\Lambda^{ab}$ forms a closed algebra and is the generator of area preserving deformation satisfying the $SO(2,1)$ Lie algebra\cite{Haldane2011}:
\begin{eqnarray}\label{apd}
[\bar\Lambda^{ab},\bar\Lambda^{cd}]=\frac{i}{2}\left(\epsilon^{ac}\bar\Lambda^{bd}+\epsilon^{ad}\bar\Lambda^{bc}+\epsilon^{bc}\Lambda^{ad}+\epsilon^{bd}\Lambda^{ac}\right)
\end{eqnarray}
Thus one can define a unitary operator
\begin{eqnarray}\label{apdg}
\hat U\left(\alpha\right)=e^{i\alpha_{ab}\bar\Lambda^{ab}}
\end{eqnarray}
such that $\hat U\left(\alpha\right)|\psi_0\rangle$ squeezes the isotropic metric $\eta_{ab}\sim\left(1,0,0,1\right)$ characterizing the isotropic quantum fluid $|\psi_0\rangle$ to a generic metric $g_{ab}$ with $\det g=1$. In the limit of small $\alpha$, the part of $\hat U\left(\alpha\right)|\psi_0\rangle$ orthogonal to $|\psi_0\rangle$ is identical to $|\psi_{|\bm q|\to 0}\rangle$.

It is thus clear that $|\psi_{|\bm q|\to 0}\rangle$ is geometric if $|\psi_0\rangle$ is translationally invariant, being equivalent to $|\psi'^{(2)}\rangle$ that is physically the quantum fluctuation of the metric from the ground state. For rotationally invariant systems with metric $\eta_{ab}$ from the ground state (note that the definition of the angular momentum requires a metric, and without loss of generality we can always take $\eta_{ab}\sim\left(1,0,0,1\right)$), we can use the commutation relations in Eq.(\ref{com1}) to define a set of ladder operators
\begin{eqnarray}\label{ladder}
\hat b_i=\frac{1}{\sqrt 2\ell_B}\left(\bar R_i^x-i\bar R_i^y\right), [\hat b_i,\hat b_i^\dagger]=1
\end{eqnarray}
The guiding center angular momentum is thus given by $\hat L_z=\frac{1}{2}\eta_{ab}\sum_i\bar R_i^a\bar R_i^b=\sum_i\hat b_i^\dagger\hat b_i+\hat b_i\hat b_i^\dagger$ commutating with the Hamiltonian. The long wavelength limit of the GMP mode is thus given by
\begin{eqnarray}
|\psi_{|\bm q|\to 0}\rangle\sim|\psi'^{(2)}\rangle=\sum_i\textsl q^2\hat b_i^2+\left(\textsl q^*\right)^2\left(\hat b_i^\dagger\right)^2|\psi_0\rangle
\end{eqnarray}
where $\textsl q=\frac{1}{\sqrt 2}\left(q_x+iq_y\right)$. Thus each electron in the ground state is boosted by a change of angular momentum of two units, and we can identify $|\psi_{|\bm q|\to 0}\rangle$ as a spin-2 excitation with respect to the ground state. Being a spin-2 excitation from the quantization of metric fluctuation is the main reason this neutral excitation is termed as the graviton excitation.

\textit{The gravitons in FQH--}It is thus important to be conceptually clear about the following points: starting with any many-body wavefunctions of the FQH quantum fluid $|\psi_0\rangle$, $|\psi'^{\left(2\right)}\rangle$ as defined above is a graviton excitation from $|\psi_0\rangle$. It should be noted that $|\psi'^{\left(2\right)}\rangle$ needs to be normalized so the magnitude $|\bm q|$ is irrelevant and can be taken to zero. On the other hand, the long wavelength GMP mode $|\psi_{|\bm q|\to 0}\rangle$ is only the graviton excitation if $|\psi_0\rangle$ is translationally invariant so that $|\psi'^{(1)}\rangle$ vanishes. 

The definition of the graviton mode is thus fundamentally independent of the GMP mode. By defining $\hat B_{m,n}=\sum_i\left(\hat b_i^\dagger\right)^m\hat b_i^n$ with $\hat B_{m,n}^\dagger=\hat B_{n,m}$, the graviton mode is given as follows:
\begin{eqnarray}\label{gdef}
|\psi_{\text{graviton}}\rangle=\frac{1}{\mathcal N}\left(e^{i\theta}\hat B_{0,2}+e^{-i\theta}\hat B_{2,0}\right)|\psi_0\rangle
\end{eqnarray}
where $\theta$ is an arbitrary phase from the direction of the unit momentum vector that we can set as zero, with the normalization given by
\begin{eqnarray}\label{gchiral}
\mathcal N^2=\langle\psi_0|\hat B_{0,2}\hat B_{2,0}+\hat B_{2,0}\hat B_{0,2}|\psi_0\rangle
\end{eqnarray}
We can clearly see the graviton mode is a linear combination of the spin $s=2$ and $s=-2$ states. This is the fundamental definition of the two ``chiral graviton" modes previously discussed in the literature\cite{Liou2019,Haldane2021,Probing2021Nguyen,wang2023geometric,balram2024splitting}, with $|\psi_{\text{graviton}}\rangle=\frac{1}{\mathcal N}\left(\mathcal N_+|\psi_g^+\rangle+\mathcal N_-|\psi_g^-\rangle\right)$ given by the two normalized states, which are clearly orthogonal, as follows:
\begin{eqnarray}\label{chiralg}
|\psi_g^+\rangle=\frac{1}{\mathcal N_+}\hat B_{2,0}|\psi_0\rangle, |\psi_g^-\rangle=\frac{1}{\mathcal N_-}\hat B_{0,2}|\psi_0\rangle
\end{eqnarray}
If the state $|\psi_0\rangle$ is translationally invariant (i.e. in the linear momentum $K=0$ sector), then $\sum_i\hat b_i|\psi_0\rangle=\sum_i\hat b_i^\dagger|\psi_0\rangle =0$, and it is easy to show that the two chiral gravitons $|\psi_g^+\rangle$ and $|\psi_g^-\rangle$ are both translationally invariant and in the $K=0$ sector. Thus in the thermodynamic limit the gravitons are also featureless in terms of the real space density distribution, carrying two units of angular momentum as compared to the ground state.

\textit{The graviton spectral sum rule--} It was first pointed out by Haldane that for any translationally invariant state $|\psi_0\rangle$ we can define a rank-2 tensor as follows\cite{Haldane2011}:
\begin{eqnarray}\label{hermitian}
\Gamma^{ab,cd}=\langle\psi_0|\bar\Lambda^{ab}\bar\Lambda^{cd}|\psi_0\rangle-\langle\psi_0|\bar\Lambda^{ab}|\psi_0\rangle\langle\psi_0|\bar\Lambda^{cd}|\psi_0\rangle
\end{eqnarray}
which can also be viewed as a four-by-four Hermitian matrix with indices $ab$ and $cd$. It is easy to see this matrix is positive semi-definite since for any metric $g_{ab}$, we have $g_{ab}\Gamma^{ab,cd}g_{cd}=\langle\psi_0|\hat L_{z,g}^2|\psi_0\rangle-\left(\langle\psi_0|\hat L_{z,g}|\psi_0\rangle\right)^2\ge 0$. Here $\hat L_{z,g}$ is the angular momentum operator defined with a general metric $g_{ab}$, while $|\psi_0\rangle$ is an eigenstate to $\hat L_{z,\eta}$.

The four-by-four matrix can be separated into its symmetric and antisymmetric part as follows. 
\begin{eqnarray}
&&\Gamma^{ab,cd}=\Gamma^{ab,cd}_S+i\Gamma^{ab,cd}_A\\
&&\Gamma^{ab,cd}_S=\frac{1}{2}\langle\{\bar\Lambda^{ab},\bar\Lambda^{cd}\}\rangle_0-\langle|\bar\Lambda^{ab}\rangle_0\langle\bar\Lambda^{cd}\rangle_0\quad\\
&&\Gamma^{ab,cd}_A=\frac{-i}{2}\langle\left[\bar\Lambda^{ab},\bar\Lambda^{cd}\right]\rangle_0\label{asym}
\end{eqnarray} 
The two parts have distinct physical meanings. The regularized static structure factor (the normalization in Eq.(\ref{graviton}) in the long wavelength is given by:
\begin{eqnarray}
\lim_{\bm q\to 0}S_{\bm q}=\frac{1}{4}\kappa\Gamma^{ab,cd}_Sq_aq_bq_cq_d
\end{eqnarray}
while the Hall viscosity or the topological shift of $|\psi_0\rangle$ is given by\cite{haldane2011self}
\begin{eqnarray}
\gamma=\eta_{ab}\langle\bar\Lambda^{ab}\rangle_0
\end{eqnarray}
which is related to $\Gamma^{ab,cd}_A$ in Eq.(\ref{asym}) via the algebra in Eq.(\ref{apd}), and without loss of generality we can always set $|\psi_0\rangle$ to be isotropic with the metric $\eta_{ab}$. It is easy to check that the Hermitian matrix of Eq.(\ref{hermitian}) has the following simple form:
\begin{eqnarray}
\left( \begin{array}{cccc}
\kappa & i\gamma & i\gamma &-\kappa \\
-i\gamma  & \kappa & \kappa & i\gamma  \\
-i\gamma  & \kappa & \kappa & i\gamma \\
-\kappa & -i\gamma & -i\gamma & \kappa \end{array} \right)
\end{eqnarray}
with the smallest eigenvalue $2\left(\kappa-\gamma\right)$ that has to be non-negative, leading to the Haldane bound\cite{haldane2009hall}
\begin{eqnarray}\label{hbound}
\kappa\ge\gamma
\end{eqnarray}
Both $\kappa$ and $\gamma$ are properties of the state $|\psi_0\rangle$ only. As long as $|\psi_0\rangle$ is translationally invariant, the Haldane bound will be universally satisfied from the algebra of the guiding center coordinates. 

\textit{The higher spin modes--} The gravitons, which are spin-2 modes, describe the simplest type of area-preserving deformation of an isotropic quantum fluid. It is the leading deformation of a circle (given by $\eta=\left(1,0,0,1\right)$), giving an ellipse described by a squeezed generic metric $g$. Geometrically we can deform the shape of the circle arbitrarily without changing its area, which is fixed by the uniform magnetic field or the magnetic length (see Fig.(\ref{fig1})). These higher order shape fluctuations go beyond just the metric fluctuations and are described by the generalized higher spin modes. 

These higher spin modes form a basis for the GMP modes going beyond the long wavelength limit. To see that, let's look at the next order of $\bm q$ expansion from Eq.(\ref{expand}), assuming translational invariance of $|\psi_0\rangle$ so the linear term vanishes:
\begin{eqnarray}
|\psi'^{\left(3\right)}\rangle=\left(-\frac{1}{2}q_aq_b\bar \Lambda^{ab}-\frac{i}{6}q_aq_bq_c\bar \Lambda^{abc}\right)|\psi_0\rangle
\end{eqnarray}
The second term is of the form $\bar \Lambda^{abc}|\psi_0\rangle\sim \left(\hat B_{3,0}+\hat B_{0,3}\right)+\left(\hat B_{1,2}+\hat B_{2,1}\right)|\psi_0\rangle$, where we ignore the coefficient of each of the terms. Thus generally speaking, the many-body wavefunctions of the (higher) spin modes (including spin-1 and spin-2 modes) are constructed as follows:
\begin{eqnarray}\label{geometric}
|\psi^{(g)}_{m,n}\rangle\sim \left(\hat B_{m,n}+\hat B_{n,m}\right)|\psi_0\rangle
\end{eqnarray}
and it is a state with spin $|m-n|$. This family of states form a complete basis describing the geometric fluctuation of the ground state, analogous to the deformation of a circle into arbitrary shapes with the same area. Indeed $\hat B_{m,n}$ generates the infinite dimensional $W_\infty$ algebra commonly used in modern string theory\cite{bergshoeff1990area,  sezgin1992area,pope1991lectures}. 

\begin{figure}
\begin{center}
\includegraphics[width=\linewidth]{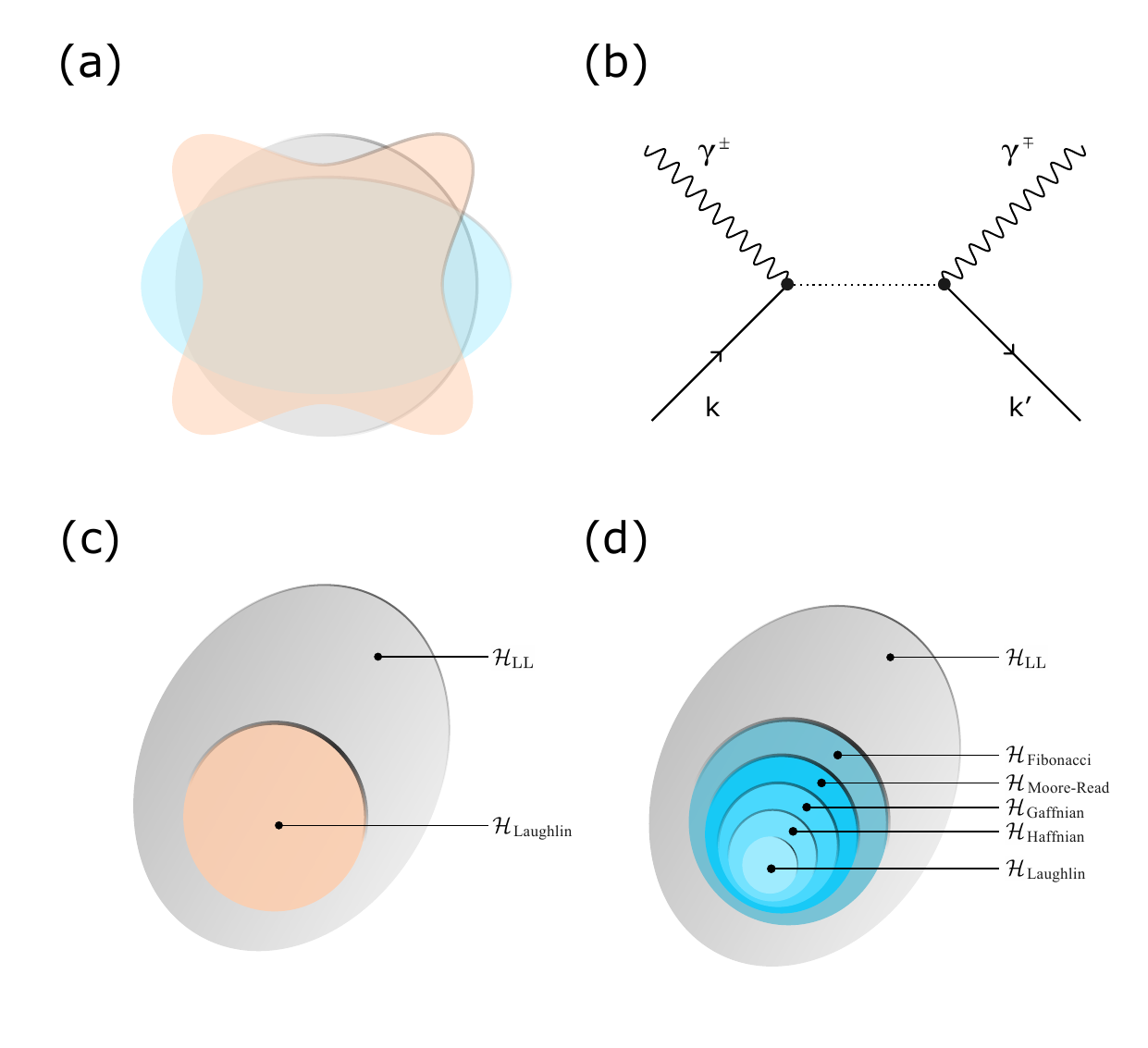}
\caption{(a). The Laughlin conformal Hilbert space $\mathcal H_{\text{Laughlin}}$ is a proper subspace within a single LL. The gravitons (or geometric modes in general) within and outside of $\mathcal H_{\text{Laughlin}}$ will have an energy difference given by the $\hat V_1$ part of the electron-electron interaction; (b). A more general hierarchy of the CHS including several phases from the Read-Rezayi series; (c) in inelastic photon scattering, the photon couples to the bare electron density non-linearly, where $\hat c^\dagger_{k'}\hat c_k$ gives the one-body operator from the density. The latter include information from all Landau levels and different bands in the 2D material; (d) the geometric fluctuation of the quantum fluid can be schematically understood as a deformation of the isotropic circle into an ellipse (orange color giving metric deformation or the gravitons) and higher order deformation (blue color giving the higher spin modes).}
\label{fig1}
\end{center}
\end{figure}

From the perspective of the SMA, the GMP modes are linear combinations of the states given by Eq.(\ref{geometric}), which are geometric fluctuations of $|\psi_0\rangle$, of which the graviton is a special case. We will refer to the states in Eq.(\ref{geometric}) as the geometric modes for the rest of this paper. In the long wavelength limit, the GMP mode is saturated by the graviton if $|\psi_0\rangle$ is translationally invariant. As the momentum increases, the higher spin modes describing higher order geometric modes become more important. These excitations are fundamentally different from the magnetoroton modes at large momenta, as the latter evolves into a quasihole-quasielectron pair. The seminal study of the higher spin modes in the Laughlin state was given in Ref.\cite{yang2014nature} and we will give a more systematic study of the higher spin modes elsewhere.

\section{Chirality of the gravitons}\label{sec_chiral}
The discussions in the previous section are completely general based on symmetry and algebra. The definition of the gravitons in Eq.(\ref{gdef}) and Eq.(\ref{gchiral}), the geometric modes in general in Eq.(\ref{geometric}), their relationship with the GMP mode from Eq.(\ref{expand}), as well as the Haldane bound in Eq.(\ref{hbound}), are formulated microscopically without the need to pick a gauge. They can also be explicitly computed on any two-dimensional manifold, be it an infinite plane or the disk, spherical or torus geometry (some subtleties will be discussed in this section). In particular, so far we have not involved the Hamiltonian in any way, which we will postpone to later sections. We thereby do not assume for now if these geometric modes can be robust and experimentally detectable; this depends on the details of the Hamiltonian. Nevertheless there are still important general statements to be made in particular about the concept of the chirality of the geometric modes.

It is clear that the geometric fluctuation itself does not have chirality; each geometric mode consists of a pair of excitations of opposite spins $m-n$ and $n-m$ given by $\hat B_{m,n}$ and $\hat B_{n,m}$. The chirality is odd under particle-hole symmetry. In the second quantized form on the infinite plane we have
\begin{eqnarray}
\hat B_{m,n}=\sum_{k=n}^\infty c^\dagger_{k+m-n}c_k\frac{\sqrt{k!\left(k+m-n\right)!}}{\left(k-n\right)!}
\end{eqnarray}
where $\hat c_k, \hat c_k^\dagger$ annihilates or creates an electron in the single particle state $|k\rangle=\left(b^\dagger\right)^k/\sqrt{k!}|0\rangle$. Defining the particle-hole conjugate operator $\hat C_{\text{ph}}$ so that $\hat C_{\text{ph}}^\dagger c_k^\dagger\hat C_{\text{ph}}=c_k$ we have the following relationship
\begin{eqnarray}
&&\hat C_{\text{ph}}^\dagger \hat B_{m,n}\hat C_{\text{ph}}=\sum_{k=n}^\infty c_{k+m-n}c^\dagger_k\frac{\sqrt{k!\left(k+m-n\right)!}}{\left(k-n\right)!}\nonumber\\
&=&-\sum_{k=m}^\infty c^\dagger_{k-m+n} c_{k}\frac{\sqrt{\left(k-m+n\right)!k!}}{\left(k-m\right)!}=-\hat B_{n,m}
\end{eqnarray}
Thus $\hat B_{m,n}$ and $\hat B_{n,m}$ are related by particile-hole symmetry, giving geometric modes of opposite chirality. 

\textit{Chiral gravitons on the sphere--} The state $|\psi_0\rangle$ upon which the operators act is generally not particle-hole symmetric. This implies the geometric modes of opposite chirality may have very different properties. It is particularly interesting to look at the chirality on different geometries of the Hall manifold. For example on the spherical geometry\cite{PhysRevLett.51.605}, geometric modes of opposite chirality are related by the inversion symmetry, so they both have the same energy and lifetime for any interaction. Thus if the gapped FQH phases (e.g. the Laughlin phase at $\nu=1/3$) are realized on the surface of the sphere, the graviton mode of both chiralities can be experimentally measured using inelastic Raman scattering with polarized photons. To see that, note on the spherical geometry the graviton modes are given by the following operators:
\begin{eqnarray}
&&\hat B_{0,2}\sim\sum_{m=-L}^LC^{2L2}_{m+2,2,m} c_{m+2}^\dagger c_m\\
&&\hat B_{2,0}\sim\sum_{m=-L}^LC^{2L2}_{m-2,-2,m} c_{m-2}^\dagger c_m
\end{eqnarray}
Thus $|\psi_g^+\rangle$ and $|\psi_g^-\rangle$ on the sphere are not only related to the particle-hole conjugate of the SMA operator, they are also related by the inversion symmetry of the sphere and are in the same $L=2$ sector. In particular $|\psi_g^+\rangle\sim \left(\hat L^+\right)^4|\psi_g^-\rangle$, where $\hat L^+=\sum_i\hat L^+_i$ with $\hat L^+_i$ raises the $L_z$ quantum number of the $i^{\text{th}}$ electron on the sphere. Thus the two chiral gravitons are localized at the opposite poles of the sphere, being the highest and the lowest weight states in the $L=2$ sector. 
\begin{figure}
\begin{center}
\includegraphics[width=\linewidth]{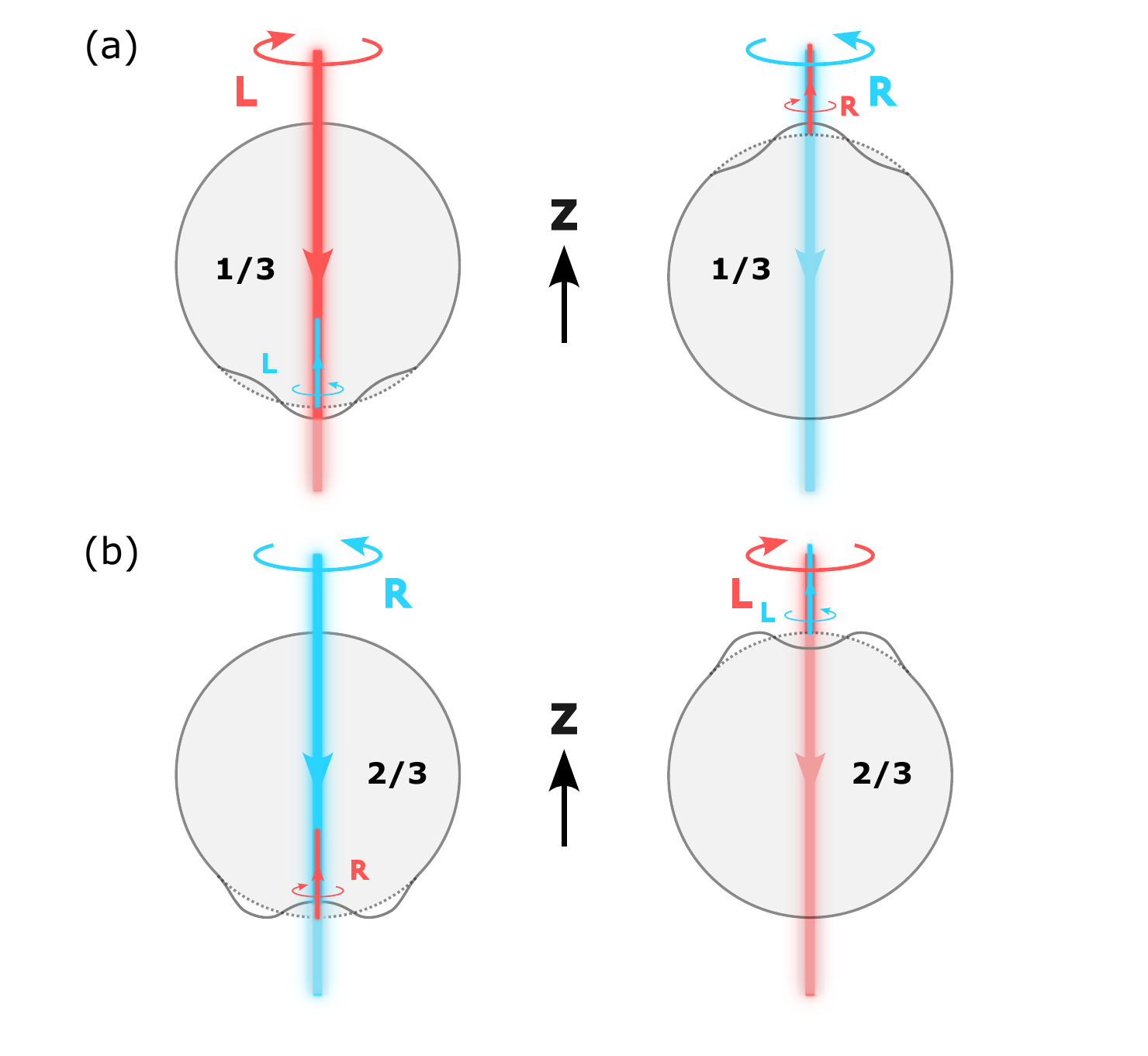}
\caption{Examples of circularly polarized light excites chiral gravitons from the ground states of the Laughlin phase on the sphere at $\nu=1/3$ and the anti-Laughlin phase at $\nu=2/3$. a). For the Laughlin state at $\nu=1/3$, an L-polarized light in the $-z$ direction is transparent at the north pole due to angular momentum conservation, but will excite a graviton at the south pole given by $\hat B_{2,0}$; the R-polarized light, however will excite the graviton at the north pole only, given by $\hat B_{0,2}$. b). For the anti-Laughlin state at $\nu=2/3$, the L-polarized light will excite the graviton at the north pole, which is induced by $\hat B_{2,0}$. Note that this operator leads to density modulation at the north pole, while for Laughlin state at $\nu=1/3$, it leads to density modulation at the south pole. The R-polarized light will excite the graviton at the south pole given by $\hat B_{0,2}$.}
\label{sphereraman}
\end{center}
\end{figure}

If we shine a circularly polarized light through the north and south pole of the sphere with the right frequency(see Fig.~\ref{sphereraman}), it will either excite $|\psi_g^+\rangle$ at the north pole, or the $|\psi_g^-\rangle$ at the south pole, as long as the graviton can be hosted by the many-body Hamiltonian (i.e. it has long enough lifetime to be detected). This is true for any polarization of the light, as for any FQH phase the gravitons of opposite chiralities come in pair and co-exist on the sphere. 

It is important to emphasize that if we only look at the incident light in the opposite direction of the magnetic field (or from outside of the sphere), as is commonly done in other geometries (e.g. disk or torus), then in many cases only one of the chiral gravitons will be excited with the proper polarization of the light, consistent with the experiments and other geometries of the Hall manifold\cite{liang2024evidence}. The proper polarization of the light depends on the nature of the state $|\psi_0\rangle$, which we will discuss in details later. Here the main message is to illustrate the consistency of Eq.(\ref{gdef}) and Eq.(\ref{gchiral}) on the spherical geometry with an interesting twist, which will also help us to understand the chiral gravitons in other geometries. It should also be noted that the argument applies to all other geometric modes with opposite chiralities. 

\textit{Chiral gravitons on the disk--} It is now easy to understand the chiral geometric modes on an infinite plane, which is equivalent to the spherical geometry with the north pole mapped to the origin, and the south pole mapped to the infinity on the plane. Thus for example when the graviton is excited from the ground state, a pair of chiral gravitons are excited, one at the origin and the other at the infinity. For a local measurement with polarized light, in cases such as the Laughlin state at $\nu=1/3$ only one chiral mode at the origin is physically relevant on an infinite plane or the thermodynamic limit. This essentially captures the main physics of the Hall sample in the experiments.

Nevertheless, the other chiral geometric mode at the infinity can be physically relevant on a disk geometry for a finite size system, as is the case in the experiment when the edge or the boundary of the quantum Hall fluid is always present. Using the Laughlin ground state $|\psi_0\rangle$ at $\nu=1/3$ as an example, the chiral graviton at the origin is given by $|\psi_g^+\rangle\sim \hat B_{0,2}|\psi_0\rangle = \sum_i \hat b_i^2|\psi_0\rangle$, which is a gapped excitation that has been experimentally measured. The other graviton with the opposite chirality is $|\psi_g^-\rangle\sim\hat B_{2,0}|\psi_0\rangle=\sum_i\left(\hat b_i^\dagger\right)^2|\psi_0\rangle$. For simplicity we take $|\psi_0\rangle$ to be the model state, so the first quantized wavefunctions are given as follows\cite{PhysRevB.87.245132}:
\begin{eqnarray}
&&|\psi_0\rangle\sim\prod_{i<j}\left(z_i-z_j\right)^3\\
&&|\psi_g^+\rangle\sim\mathcal A\left[\left(z_1-z_2\right)\prod'_{i<j}\left(z_i-z_j\right)^3\right]\\
&&|\psi_g^-\rangle\sim\prod_iz_i^2\prod_{j<k}\left(z_i-z_j\right)^3\label{g3}
\end{eqnarray}
Here we omitted the irrelevant Gaussian part of the wavefunction; $\mathcal A$ indicates antisymmetrization over all particle indices, and the prime sign on $\prod'_{i<j}$ indicates the product of pairs of $\{i, j\}$ that is not $\{1,2\}$. The physics is the same if $|\psi_0\rangle$ is the more realistic ground state of the Coulomb interaction, just that we do not have nice first quantized wavefunctions. One can immediately see from Eq.(\ref{g3}) that $|\psi_g^-\rangle$ is an edge excitation on the disk that is gapless: the other chiral graviton is a gapless density modulation at the edge, physically related to the shape deformation of the entire quantum Hall droplet as it should be. In realistic systems, such low energy neutral excitations can in principle be measured, with the chirality of such excitations opposite to the chirality of the gapped graviton at high energy. 
\begin{figure}
\begin{center}
\includegraphics[width=\linewidth]{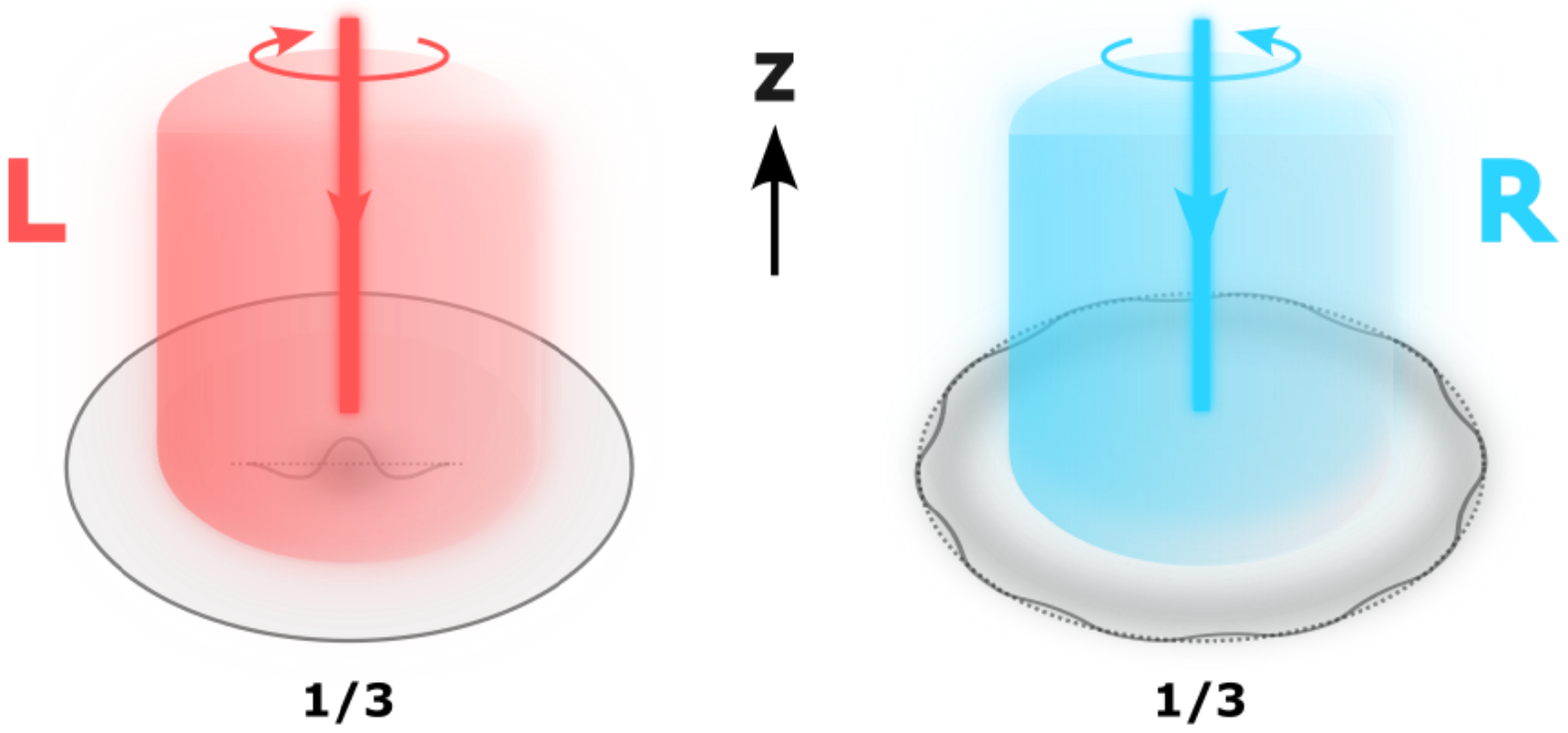}
\caption{The schematics of circularly polarized light exciting chiral gravitons from the FQH ground states on the disk geometry. a) The excitation of the bulk graviton mode, with density modulation in the bulk. b) The excitation of the edge graviton mode of the opposite chirality, with density modulation confined to the edge.}
\label{diskraman}
\end{center}
\end{figure}

Thus for a Hall manifold with a boundary, the two chiral gravitons can be understood as the bulk and edge gravitons. They have the opposite chiralities and while the bulk graviton is gapped at high energy, the gapless edge graviton has low energy that depends on the confining potential at the edge. For the Laughlin state at $\nu=1/3$, $\psi_g^+\rangle$ is the bulk graviton and $|\psi_g^-\rangle$ is the edge graviton. For the particle-hole conjugate anti-Laughlin state at $\nu=2/3$, the bulk graviton is $|\psi_g^-\rangle$ while the edge graviton is $|\psi_g^+\rangle$. Determining which chiral graviton is the bulk graviton depends on the density modulation of the many-body wavefunction, which we will discuss in more details in Sec.~\ref{sec_exp}.

 \textit{Chiral gravitons on the torus--} Unlike other geometries (e.g. sphere, disk and infinite plane), the torus geometry does not have rotational invariance. Rotational symmetry is locally restored for large system sizes, thus for excitations that are localized (e.g quasiholes or quasielectrons) we have rotational invariance as long as the system is large enough. For states that are not localized, for example the FQH ground state or the geometric modes (including the gravitons), the periodic boundary condition breaks rotational symmetry even in the thermodynamic limit. One can see that from a translationally invariant state $|\psi_0\rangle$, which will be annihilated by the guiding center coordinates, $\bar R_x|\psi_0\rangle=\bar R_y|\psi_0\rangle = 0$ on an infinite plane. In contrast, it is difficult to define such translation generators on the torus, that can annihilate a uniform quantum state on the torus in the $(K_x, K_y) = (0, 0)$ sector. Here $K_x, K_y$ are the crystal momentum on the torus. 

Thus the use of $\hat B_{m,n}$ to generate the geometric modes is quite subtle, given that $\hat B_{m,n}$ does not satisfy the periodic boundary condition on the torus. The guiding center density operator $\bar\rho_{\bm q}$, on the other hand, is always well defined as the translation operator, as long as $\bm q$ is one of the allowed crystal momentum on the torus. In the thermodynamic limit we can thus define the following operators:
\begin{eqnarray}
\hat T_{\theta,m}=\lim_{\bm q\to 0}\frac{1}{|\bm q|^{m}}\left(\bar\rho_{\bm q}-\langle\psi_0|\bar\rho_{\bm q}|\psi_0\rangle\right)
\end{eqnarray}
and the geometric operator $\hat B_{m,n}$ is given by linear combinations of $\hat T_{\theta,m}$, where $\theta$ gives the direction of $\bm q$. For example the chiral graviton operators are given by
\begin{eqnarray}
&&\hat B_{2,0}\sim \hat T_{0,2}-\hat T_{\pi/2,2}+\frac{i}{2}\left(\hat T_{\pi/4,2}-\hat T_{3\pi/4,2}\right)\\
&&\hat B_{0,2}\sim \hat T_{0,2}-\hat T_{\pi/2,2}-\frac{i}{2}\left(\hat T_{\pi/4,2}-\hat T_{3\pi/4,2}\right)
\end{eqnarray}

Given that the limit of $\bm q\to 0$ can only be taken in the thermodynamic limit, for finite systems we can only obtain the GMP modes using $\hat T_{\theta,m}$ which is a linear combination of all the geometric modes, though increasingly dominated by the gravitons for larger system sizes. An alternative way of constructing the graviton modes on the torus geometry is to use a two-body operator introduced in a series of seminal papers in Ref.\cite{Liou2019,yang2016acoustic}. We will discuss about that approach in more details in Sec.~\ref{sec_connection}.

\section{Dynamics of geometric fluctuations}\label{sec_dynamics}
For these geometric modes to be physically realized and even experimentally measured, it depends on both the variational energy of the geometric mode and its lifetime. A good measure of analyzing both the energy and the lifetime is to look at the spectral function defined as follows:
\begin{eqnarray}\label{spectral}
I\left(E\right)=\sum_{n>0}\langle\psi_{m,n}^{(g)}|\psi_n\rangle\langle\psi_n|\psi_{m,n}^{(g)}\rangle\delta\left(E-E_n\right)
\end{eqnarray}
where $|\psi_n\rangle$ are the excitation eigenstates and $E_n$ are the energies, or the eigenvalues of the Hamiltonian.

The computation of the spectral function still needs to be done numerically. So far we do not have good arguments on whether or not the graviton modes, or the geometric modes in general, are long lifetime excitations with either model or realistic interaction, especially since these modes are generally buried in the continuum of gapped excitations. There are however a number of things we can determine based on the general properties of the Hamiltonians, and we will leave detailed numerical computations for future works.

\textit{Multiple gravitons--} In the previous section we have clarified the concept of chiralities of the geometric modes, and in particular for the graviton modes. In this section, we discuss the concept of ``multiple gravitons" that was proposed in the literature, from the microscopic formulation we have so far established. The generalization to other geometric modes is quite straightforward.

To conceptually understand what ``multiple gravitons" fundamentally are, it is useful to start by not assuming an infinite magnetic field, so we do not confine ourselves to a single LL as we have done in the previous sections. It is the bare electron density $\rho_{\bm q}=\sum_ie^{iq_a\left(\tilde R^a+\bar R^a\right)}$ that couples to the photons in the experiment, and the proper definition of the neutral excitation as the graviton should come from the real space metric deformation generated by $\hat\Lambda^{ab}=\tilde\Lambda^{ab}+\bar\Lambda^{ab}$, where analogously we have $\tilde\Lambda^{ab}=\sum_i\tilde R_i^a\tilde R_i^b-\langle\psi_0|\sum_i\tilde R_i^a\tilde R_i^b|\psi_0\rangle$, again the upper indices are implicitly symmetrized. The proper graviton mode should thus be defined by
\begin{eqnarray}
|\tilde\psi_g\rangle\sim\lim_{\bm q\to 0}q_aq_b\hat\Lambda^{ab}|\psi_0\rangle
\end{eqnarray}
Without loss of generality we take $|\psi_0\rangle$ to be within the LLL. Note that even if $|\psi_0\rangle$ is translationally invariant (e.g. the ground state of the integer quantum Hall effect), the long wavelength limit of $\rho_q$ is not the graviton, since the cyclotron part linear in $\tilde R^a$ does not annihilate $|\psi_0\rangle$. Nevertheless the graviton mode is a well-defined state, involving the mixing of the LLL and the second Landau level (2LL) with an energy gap of twice the cyclotron energy. Treating $\hat\Lambda^{ab}$ as the graviton operator, we can thus rewrite it in a slightly more illuminating way:
\begin{eqnarray}
&&\hat\Lambda^{ab}=\Lambda^{ab}_{LL}+\Lambda'^{ab}_{LLL}\\
&&\Lambda'^{ab}_{LLL}=\hat P_{LLL}\hat\Lambda^{ab}\hat P_{LLL}
\end{eqnarray} 
where $\hat P_{LLL}$ is the projection into the LLL, and in this case obviously $\Lambda^{ab}_{LL}=\tilde\Lambda^{ab}, \Lambda'^{ab}_{LLL}=\bar\Lambda^{ab}$. The Hamiltonian of the system can also be written in the following way:
\begin{eqnarray}
\hat H=\hat P_{LLL}+\lambda\hat H_{int}
\end{eqnarray}
where the first term is equivalent to the kinetic energy, and in the limit of large magnetic field it dominates the second interaction term with $\lambda\ll 1$. We can then define the following graviton modes:
\begin{eqnarray}
&&|\psi_{LL,g}\rangle\sim\Lambda^{ab}_{LL}|\psi_0\rangle\\
&&|\psi_{LLL,g}\rangle\sim\Lambda'^{ab}_{LLL}|\psi_0\rangle
\end{eqnarray}
Since we have $\hat P_{LLL}|\psi_0\rangle=|\psi_0\rangle$, the energy gap of $|\psi_{LL,g}\rangle$ will be of order unity (the order of the cyclotron energy), while that of $|\psi_{LLL,g}\rangle$ is of the order $\lambda$. Note that by definition we have $\langle \psi_{LL,g}|\psi_{LLL,g}\rangle=0$. Here we can consider $|\psi_{LL,g}\rangle$ and $|\psi_{LLL,g}\rangle$ as two graviton modes: the \emph{cyclotron graviton} that mixes different LLs, and the \emph{LL graviton} that stays within a single LL. This establishes the concept of graviton multiplicity for all other cases both at a formal and intuitive level.

We now focus on the interaction Hamiltonian, and the realistic model for the recent experiments is the LLL Coulomb interaction. To illustrate the essential physics cleanly, we choose $\hat H_{\text{int}}=\hat V_1+\lambda'/\lambda\delta V$, where $\hat V_1$ is the pseudopotential model Hamiltonian\cite{PhysRevLett.51.605} for the Laughlin phase at $\nu=1/3$. This is a two-body interaction describing the shortest range repulsion between fermions. Thus we can rewrite the pseudopotential Hamiltonian in the equivalent form $\hat V_1=\sum_{i<j}\hat P_{ij,1}$, where $\hat P_{ij,m}$ is the projection operator for a pair of electrons $i$ and $j$ with relative angular momentum $m$. More importantly, $\hat H_{int}$ defines a null space $\mathcal H_{\text{Laughlin}}$ such that for any many-body state $|\psi\rangle\in\mathcal H_{\text{Laughlin}}$ we have $\hat V_1|\psi\rangle=0$. Thus $\mathcal H_{\text{Laughlin}}$ is a well-defined Hilbert space spanned by the model Laughlin ground states and quasihole states (see Fig.(\ref{fig1})). 

$\mathcal H_{\text{Laughlin}}$ is named as the Laughlin conformal Hilbert space\cite{yang2021gaffnian,yang2022anyons, yang2022composite, trung2023spin, wang2021analytic, wang2023geometric}, which for now is just a name but we will explain the rationale in Sec.~\ref{sec_chs}. We can now define a projection operator $\hat P_{\text{Laughlin}}$ analogous to $\hat P_{LLL}$, which projects into $\mathcal H_{\text{Laughlin}}$, a proper subspace of the Hilbert space of a single Landau level. It is then useful to note that $\hat V_1=\hat P_{\text{Laughlin}}$, so the full Hamiltonian can be expressed as follows:
\begin{eqnarray}\label{ppham}
\hat H=\hat P_{LLL}+\lambda\hat P_{\text{Laughlin}}+\lambda'\delta V
\end{eqnarray}
Here we also take $\lambda'\ll\lambda$, and we would like to emphasize here that Eq.(\ref{ppham}) is adiabatically connected to the LLL Coulomb interaction for several experimentally relevant topological phases, including the Laughlin $\nu=1/3$ and anti-Laughlin $\nu=2/3$ phases, the Jain $\nu=2/5$ and $\nu=2/7$ phases, as well as the conjectured composite Fermi liquid (CFL) phase at $\nu=1/4$.

The graviton operator can now be expressed as:
\begin{eqnarray}
&&\hat\Lambda^{ab}=\Lambda^{ab}_{LL}+\Lambda^{ab}_{LLL}+\Lambda^{ab}_{\text{Laughlin}}\\
&&\Lambda_{\text{Laughlin}}=\hat P_{\text{Laughlin}}\Lambda'^{ab}_{LLL}\hat P_{\text{Laughlin}}
\end{eqnarray}
with which we can define three graviton modes as
\begin{eqnarray}
&&|\psi_{LL,g}\rangle\sim\Lambda^{ab}_{LL}|\psi_0\rangle\\
&&|\psi_{LLL,g}\rangle\sim\Lambda^{ab}_{LLL}|\psi_0\rangle\\
&&|\psi_{\text{Laughlin,g}}\rangle\sim\Lambda^{ab}_{\text{Laughlin}}|\psi_0\rangle
\end{eqnarray}
Given that it is completely within the Laughlin conformal Hilbert space $\mathcal H_{\text{Laughlin}}$, we call $|\psi_{\text{Laughlin,g}}\rangle$ the Laughlin graviton. For $|\psi_0\rangle$ within the LLL, the cyclotron graviton is always present. Since it is at very high energy (on the order of the cyclotron gap), and also for reasons we will explain later, we are not very interested in it in this work. We include the discussions about the cyclotron graviton here to illustrate that the concept of multiple gravitons is not ``new physics", but rather inherent in the basic formulation of the geometric fluctuations of the quantum Hall fluids.

For the rest of the paper we are going to ignore the cyclotron graviton and see how the multiplicity of the gravitons appear analogously within a single LL. It is important to know that $\Lambda^{ab}_{LLL}$ and $\Lambda^{ab}_{\text{Laughlin}}$ depend on the nature of the many-body state $|\psi_0\rangle$ due to the null space projection $\hat P_{\text{Laughlin}}$. If $|\psi_0\rangle$ is outside of $\mathcal H_{\text{Laughlin}}$, then clearly it is annihilated by $\Lambda^{ab}_{\text{Laughlin}}$ and the Laughlin graviton does not exist. This is the case for the ground state of FQH phases at the filling factor $\nu>1/3$, such as the Jain state at $\nu=2/5$, the Moore Read state at $\nu=1/2$ and the anti-Laughlin state at $\nu=2/3$. For the Laughlin state at $\nu=1/3$, $\Lambda'^{ab}_{LLL}|\psi_0\rangle$ is completely outside of $\mathcal H_{\text{Laughlin}}$, so the Laughlin graviton also does not exist.

For the quantum Hall fluids at filling factor $\nu<1/3$, in general $\Lambda^{ab}_{\text{Laughlin}}|\psi_0\rangle$ does not vanish. For FQH states at $\nu=2/7$ and the CFL at $\nu=1/4$, both the LL graviton and the Laughlin graviton are present, though for CFL the Laughlin graviton may be gapless due to the gapless nature of the CFL\cite{jain2015composite}. These are the cases of multiple gravitons discussed in the literature\cite{Nguyen2021, balram2021highenergy, wang2023geometric}. However it is important to note that the ``new graviton" here is actually the Laughlin graviton at low energy, instead of the high energy LL graviton. The previous identification of the LL graviton as the additional ``Haldane mode" in the effective field theory is not appropriate, since the LL graviton is universally present at the same energy for FQH phases with $\nu\ge 1/3$.

Not all FQH phases at $\nu<1/3$ have the LL graviton, and thus two graviton modes. If $\Lambda'^{ab}_{LLL}|\psi_0\rangle\in\mathcal H_{\text{Laughlin}}$, we then have $\Lambda'^{ab}_{LLL}=\Lambda^{ab}_{\text{Laughlin}}$ and thus $\Lambda^{ab}_{LLL}=0$. The simplest examples are the Laughlin states $\nu=1/m$ with $m>3$, and one can analytically show\cite{wang2021analytic} that $\Lambda^{ab}_{LLL}=0$ and thus only the Laughlin graviton is present at the low energy (at the energy scale of $\lambda'$). Therefore with respect to the Hamiltonian Eq.(\ref{ppham}), the number of graviton modes within a single LL for any quantum Hall fluids, whether or not it is gapped or gapless, can be qualitatively determined. The energy scales of the graviton gaps can also be easily estimated from the Hamiltonian and the incompressibility gap of the quantum Hall fluids. All the arguments and conclusions from above also apply for the realistic Coulomb interaction within the LLL.

It is useful to go over the nomenclature of the gravitons here and what have been used in the literature, and more details will be discussed in Sec.~\ref{sec_connection}. The LL graviton is the only graviton of the Laughlin state at $\nu=1/3$, the first discovered graviton. It is also the graviton of the Moore-Read state, the Jain states, etc, and also the high energy graviton of the FQH phases with $\nu<1/3$, such as the states at $\nu=2/7$ and $\nu=2/9$. It is also the only gapped graviton of the CFL state at $\nu=1/4$. These gravitons are all called the LL graviton in this work, because physically they all come from the guiding center metric fluctuation within a single LL but outside of $\mathcal H_{\text{Laughlin}}$, thus occuring at similar variational energy, being punished by the $\hat V_1$ part of the Coulomb interaction. The LL graviton was initially mis-identified as the additional ``Haldane mode"\cite{Nguyen2021}; in the CF/parton theory it was also named the ``parton graviton"\cite{balram2021highenergy}, though in principle the parton construction is not needed for this graviton for the Laughlin phase.

The Laughlin graviton in contrast is within $\mathcal H_{\text{Laughiln}}$ and is thus not punished by the $\hat V_1$ part of the Coulomb interaction. It is thus the additional graviton mode for $\nu<1/3$ generally occurring at lower energy as compared to the LL graviton. It is more appropriate to identify the Laughlin graviton as the additional ``Haldane mode" needed in the effective field theory to satisfy the Haldane bound\cite{haldane2009hall, nguyen2018particle, kumar2024numerical}. In the CF/parton theory it is simply called the CF graviton as it does not involve the Jastrow factors describing the flux attachment in the CF theory.

\section{Geometric fluctuations of conformal Hilbert spaces}\label{sec_chs}

The two projection operators in Eq.(\ref{ppham}) define two Hilbert spaces for the quantum Hall fluids: $\hat P_{LLL}|\psi\rangle\in\mathcal H_{LLL}$ and $\hat P_{\text{Laughlin}}|\psi\rangle\in\mathcal H_{\text{Laughlin}}$. Here $\mathcal H_{LLL}$ is the Hilbert space of the LLL, corresponding to the ground state and the hole states of the integer quantum Hall (IQH) phase; $\mathcal H_{\text{Laughlin}}$ is the Hilbert space of the Laughlin phase, spanned by its ground states and the quasihole states. These two are examples of conformal Hilbert space (CHS) with emergent conformal symmetry, capturing the low energy universal topological physics of the respective FQH phases\cite{yang2021gaffnian,yang2022anyons, yang2022composite, trung2023spin, wang2021analytic, wang2023geometric}. The two gravitons defined by $\hat P_{LLL}$ and $\hat P_{\text{Laughlin}}$ thus also have explicit geometric meaning: they are neutral excitations from the quantum fluctuations of the metrics characterizing the respective CHS. This rigorous interpretation can be extended to all geometric modes. For each geometric mode defined by $\hat B_{m,n}+\hat B_{n,m}$, it in principle consists of multiple modes defined by the projection operators, corresponding to the more general quantum geometric fluctuations of the respective CHS.

Given the richness of the FQH fluids, there is a hierarchy of CHS with a single LL or $\mathcal H_{LLL}$ (see Fig.(\ref{fig1})) which was discussed extensively in Ref.\cite{wang2023geometric}. A typical example is the Read-Rezayi series giving a well-defined hierarchy of CHS, each with model Hamiltonians and the hierarchy is given by $\mathcal H_{\text{Laughlin}}\in\mathcal H_{\text{Moore-Read}}\in\mathcal H_{\text{Fibonnaci}}\in\cdots\in\mathcal H_{LLL}$. The two additional CHS's are spanned by the ground states and the quasihole states of the three-body Moore-Read and four-body Fibonacci Hamiltonians respectively. In general given the following hierarchy of the CHS
\begin{eqnarray}
\mathcal H_1\in\mathcal H_2\in\mathcal H_3\cdots\in\mathcal H_n
\end{eqnarray}
we can define a series of corresponding projection operators $\hat P_k$ that projects into the Hilbert space of $\mathcal H_k$. These operators allow us to define a series of graviton operators $\Lambda^{ab}_{k}=\hat P_{k+1}\bar \Lambda^{ab}\hat P_{k+1}-\hat P_{k}\bar\Lambda^{ab}\hat P_{k}$, with $\hat P_0$ projecting into the vacuum. We thus have
\begin{eqnarray}\label{chsapd}
\bar\Lambda^{ab}=\sum_{k=0}^{n-1}\Lambda^{ab}_k
\end{eqnarray}
Each $\Lambda^{ab}_k$ projects into $\mathcal H_k\backslash \mathcal H_{k-1}$ with $\mathcal H_0$ being the null space, physically describing the quantum metric fluctuation of $\mathcal H_k$; the corresponding graviton mode is thus given by $\Lambda^{ab}_k|\psi_0\rangle$.

 \textit{The number of graviton modes--} There are in principle an infinite number of hierarchical CHS's within a single LL, for example from the Read-Rezayi series\cite{read1999beyond}, or from the composite fermion picture\cite{jain1989composite,yang2022composite}. As one can see from Eq.({\ref{chsapd}), an infinitesimal metric deformation generated by $\bar\Lambda^{ab}$ leads to the same metric deformation of all CHS's. The only reason for us to distinguish between different ``graviton modes" generated by different $\Lambda_k^{ab}$ comes from the energetics: given a particular Hamiltonian, the energy cost of metric deformation for different CHS may be different. Thus the notion of the number of ``graviton modes" is only meaningful with respect to a particular Hamiltonian, of which Eq.(\ref{ppham}) is an example. Generally speaking if the realistic Hamiltonian is a good approximation of the following model Hamiltonian
 \begin{eqnarray}\label{modelpham}
 \hat H=\sum_{k'=0}^{n'}\lambda_{k'}\hat P_{k'}
 \end{eqnarray}
 with $\hat P_{k'}$ projects into a CHS $\mathcal H_{k'}$ with $\mathcal H_{k_1}\in\mathcal H_{k_2}$ and $\lambda_{k_1}\ll\lambda_{k_2}$ if $k_1<k_2$, then we will have at most $n'$ graviton modes with well separated spectral peaks in the spectral function given by Eq.(\ref{spectral}). The Haldane bound, on the other hand, is always satisfied with or without Eq.(\ref{modelpham}) and its physics is not really relevant to the concept of multiple gravitons. 
 
The Hamiltonian of Eq.(\ref{modelpham}) is by construction the model Hamiltonian for all FQH phases defined by each of $\hat P_k$, as long as we diagonalise it at different filling factors. It is not explicitly a local Hamiltonian, but in most if not all cases (maybe with some exceptions of the CF states with no exact local Hamiltonians) it is equivalent to a local model Hamiltonian as one can see from the construction of Eq.(\ref{ppham}). Let $|\psi_{0, k}\rangle$ be the FQH ground state corresponding to the CHS of $\mathcal H_k$, then obviously it is going to be annihilated by all $\Lambda^{ab}_{k'}$ with $k'<k$. Since $|\psi_{0,k}\rangle$ is the unique highest density state within $\mathcal H_k$, then all neutral excitations are outside of $\mathcal H_k$, so $\Lambda^{ab}_k$ will also annihilate $|\psi_{0,k}\rangle$, but $\Lambda^{ab}_{k+1}|\psi_{0,k}\rangle$ will always be a non-vanishing graviton mode.

The question is if $\Lambda^{ab}_{k'}$ with $k'>k+1$ will annihilate $|\psi_{0,k}\rangle$ so that there can be more than one graviton modes. This crucially depends on the nature of $|\psi_{0,k}\rangle$ as we have discussed extensively in Sec.~\ref{sec_dynamics}. Generally speaking, one can show either analytically or from strong numerical evidence that $\bar\Lambda^{ab}|\psi_{0,k}\rangle\in\mathcal H_p$ with $p\ge k+1$, thus $\Lambda^{ab}_{k'}$ with $k'>p$ will also annihilate $|\psi_{0,k}\rangle$, and with respect to Eq.(\ref{modelpham}) or all realistic Hamiltonians well approximated by Eq.(\ref{modelpham}), there are in total $p-k$ graviton modes (or more precisely the spectral peaks) for the FQH phase of $|\psi_{0,k}\rangle$.

\section{Experimental measurement of the geometric modes}\label{sec_exp}

In this section we will discuss about the experimental measurements of the geometric modes in general, instead of singling out the gravitons as special cases. As we have shown from the previous sections, if a realistic Hamiltonian in the experiment can be well approximated by a model Hamiltonian in the form of Eq.(\ref{modelpham}), we can then define a family of model wavefunctions as follows:
\begin{eqnarray}
&&|\psi_{m,n,k}^+\rangle\sim\hat B_{m,n,k}|\psi_0\rangle,\quad |\psi_{m,n,k}^-\rangle\sim\hat B_{n,m,k}|\psi_0\rangle\quad\\
&&\hat B_{m,n,k}=\hat P_{k+1}\hat B_{m,n}\hat P_{k+1}-\hat P_{k}\hat B_{m,n}\hat P_{k}
\end{eqnarray}
where $|\psi_{m,n,k}^{\pm}\rangle$ is a neutral excitation with chirality $\pm$ describing the geometric fluctuation of the ground state quantum fluid $|\psi_0\rangle$ associated with the CHS $\mathcal H_k$, with $|\psi_0\rangle\in\mathcal H_k$. 

Depending on the nature of the ground state, many of the geometric modes above do not exist as we have explained in the previous section. For the LLL Coulomb interaction, the relevant model Hamiltonian is Eq.(\ref{ppham}) so that for each pair of non-negative integers $\{m,n\}$, there are at most two geometric modes (or four if the two chiralities are counted). One geometric mode is outside of the Laughlin conformal Hilbert space $\mathcal H_{\text{Laughlin}}$, thus punished by $\hat V_1$ or the shortest range part of the interaction. The other mode is inside $\mathcal H_{\text{Laughlin}}$ and thus has the energy from the long range tail of the interaction $\delta V$.

All these non-vanishing geometric modes are well-defined neutral excitations in the long wavelength limit. Whether or not these neutral excitations can be observed in the experiment depends on their lifetime, since they are generally speaking not the exact eigenstates of the microscopic Hamiltonian. Numerically the predictions can be made by computing the spectral function Eq.(\ref{spectral}) for each of the geometric mode, and the sharpness of the resonant peaks quantitatively give the lifetime and thus the ease of being experimentally observed. From this perspective, such numerical computations already can give information on which chiral geometric modes can be experimentally observed, for example when coupling to the photons non-linearly.
 
 \textit{Chiral geometric modes in experiments--} As we have shown analytically, for any quantum Hall fluid the geometric modes of both chiralities are present. On the spherical geometry modes of the opposite chiralities are related by inversion symmetry and thus have the same energy and lifetime. Before talking about the more realistic case of the plane or disk geometry, it is interesting to think about which chiral mode can be measured by the incoming photons through the north and south pole on the sphere. For example with the Laughlin ground state at $\nu=1/3$, a right-handed circularly polarized light will be scattered by the $|\psi^+\rangle$ at the north pole, but will pass through transparently at the south pole where $|\psi^-\rangle$ is located. In contrast a left-landed circularly polarized light will be transparent at the north pole but will scatter at the south pole (see Fig.(\ref{sphereraman})). The situation will be the opposite if we do the experiment on the anti-Laughlin ground state at $\nu=2/3$. For gravitons, these predictions can be made based on the numerical computations of the spectral functions done in the literature.
 
In fact for gravitons and all other geometric modes, such predictions can be made by just looking at the electron density modulation of the neutral excitations, without the need of expensive computation of the spectral function requiring the knowledge of the entire energy spectrum. While on a torus geometry or an infinite plane, all geometric modes are translationally invariant with uniform density distribution, this is no longer the case for a finite spherical geometry. Due to the non-zero Gaussian curvature of the finite sphere, geometric modes are neutral excitations of intrinsic spin that can have non-uniform density distribution. For example, for the highest weight or lowest weight geometric modes the electron density modulations are concentrated at the north or south pole. For each mode, the electron density relaxes back to the ground state density towards the other pole, with vanishing density modulation.

The external probe, such as the circularly polarized light, only couples to the density modulation. For the Laughlin $\nu=1/3$ state, the chiral geometric modes generated by $\hat B_{m,n}, n>m$ host density modulation at the north pole. Thus as far as the chiral graviton is concerned, only the north pole will couple to the right-handed circularly polarized light, though due to the angular momentum conservation, it is transparent to the left-handed circularly polarized light. The other generator $\hat B_{m,n}, m<n$ leads to only the density modulation at the south pole, so if we probe with the incident light from outside of the sphere at the north pole, this chiral mode will not be detected.

Due to particle-hole symmetry, for the anti-Laughlin state at $\nu=2/3$ we have $\hat B_{m,n},n>m$ that generates the density modulation at the south pole instead of the north pole. Thus due to angular momentum conservation, a local measurement at the north pole with the incident light from outside of the sphere can only couple to the graviton mode with the left-handed circularly polarized light. 

This simple analysis can be easily generalized to each of the spectral peak if there are multiple of them (i.e. the multiplicity of the geometric modes), describing the geometric fluctuations of each CHS as the subspace within a single LL. For example it has been found that the composite fermion state at $\nu=2/7$ has two graviton modes with opposite chiralities\cite{nguyen2021multiple, wang2023geometric}. This is because each of the $\hat B_{m,n}$ (and $\hat B_{0,2}$ in particular for the gravitons) induces density modulations both at the north and the south pole, in contrast to states like Laughlin $\nu=1/3$ where each $\hat B_{m,n}$ only induces density modulation at one of the poles. More precisely following the analysis in Sec.~\ref{sec_dynamics}, $\hat B_{m,n}$ projected into $\mathcal H_{\text{Laughlin}}$ only induces density modulation at the south pole for $m<n$; in contrast $\hat B_{m,n}$ outside of $\mathcal H_{\text{Laughlin}}$ induces density modulation at the north pole for $m<n$. The $\nu=2/7$ comes from the anti-Laughlin state of the composite fermions\cite{yang2022composite} at filling factor $\nu=2/3$ (not the electron filling), and each composite fermion consists of two fluxes attached to one electron. This particle-hole conjugation of the composite fermions leads to the opposite chirality of the graviton within $\mathcal H_{\text{Laughlin}}$.

 \textit{Challenges of measuring geometric modes--} Even though the geometric modes of the quantum Hall fluids are well defined many-body states, experimental measurement of such states requires them to have sharp spectral peaks with respect to the realistic Hamiltonians. The multiple spectral peaks for the geometric modes are also only possible if the Hamiltonian approximates some form of Eq.(\ref{modelpham}). If we focus on inelastic Raman scattering with circularly polarized photons, the important question is if there are other constraints of observing these geometric modes, even if they are present with sharp spectral peaks.
 
 As one can see the coupling of the photons to the electrons in the many-body quantum fluid is fundamentally via the density modes of the quantum fluids. The reason for gravitons to be experimentally relevant in this context is that in the limit of small momentum transfer (ideally suited for photon coupling), the density modes are the graviton modes \emph{only if} the ground state is translationally invariant (see Sec.\ref{sec_density}). If the ground state is not translationally invariant (e.g charge-density wave or stripe phase) and is thus non-degenerate, the long wavelength limit of the density mode is saturated by the center-of-the-mass excitation within the degenerate manifold. We are thus not able to detect gravitons via Raman scattering for such quantum Hall fluids. The same argument also applies to the cyclotron gravitons: since the linear part of the cyclotron density operator does not vanish even if the ground state is translationally invariant (e.g. the IQH ground state), the long wavelength linear response is saturated by the $s=1$ excitons from the mixing with the first Landau level (while for the $s=2$ graviton it is the mixing with the second Landau level). This makes it difficult for the cyclotron graviton to be detected in experiment.

Thus the experimental detection of the gravitons in the long wavelength limit can be used as a useful indication that the ground state is translationally invariant. These include of course the gapped FQH topological phases and the well studied CFL at $\nu=1/4$. The absence of the gravitons in experiments at other generic filling can be used as strong evidence for the compressible bubble or stripe phases for the underlying quantum Hall fluids.
 
From the order by order expansion of the density operator, we can see that even if we can couple to the density modes of the ground state (translationally invariant or not) with large momentum transfer, the response will be a mixture of all geometric modes dominated by lower spin modes. Thus if we can couple to these modes with the right angular momentum transfer, higher spin modes can be experimentally distinguished but their signals will be much weaker as compared to, for example, the graviton modes. 
 
 \textit{Graviton(s) of compressible quantum Hall fluids--}The formulation and the universal properties of the geometric modes do not depend on whether the quantum Hall fluid is compressible or not. However if the quantum Hall fluid is compressible, the neutral excitations can become gapless, including the geometric modes or more specifically the gravitons. This however concerns the non-universal dynamic properties of the geometric mode determined by the details of the Hamiltonian. From the theory of the hierarchy of the CHS, it is easy to see that certain geometric modes will have finite energy as reflected from the spectral function, even if the ground state quantum Hall fluid is compressible.
 
Let us look at the gravitons as an example with respect to the LLL Coulomb interaction at filling factor $\nu<1/3$ and the ground state is within $\mathcal H_{\text{Laughlin}}$. Without loss of generality we can understand the physics with the adiabatically connected model Hamiltonian Eq.(\ref{ppham}). Assuming the graviton state $\bar\Lambda^{ab}|\psi_0\rangle$ is not entirely within $\mathcal H_{\text{Laughlin}}$, both $\Lambda^{ab}_{LLL}|\psi_0\rangle$ and $\Lambda^{ab}_{\text{Laughlin}}|\psi_0\rangle$ will not vanish. This is generally the case for the filling factor $\nu>1/5$, including for example the incompressible phase at $\nu=2/7$ and the compressible composite Fermi liquid at $\nu=1/4$. It is easy to see that the variational energy of $\Lambda^{ab}_{LLL}|\psi_0\rangle$ is always on the order of $\lambda$, which is the energy scale of the short range $\hat V_1$ interaction. Thus if the incompressibility gap of the quantum fluid is determined by $\delta V$ within $\mathcal H_{\text{Laughlin}}$, then only the energy of $\Lambda^{ab}_{\text{Laughlin}}|\psi_0\rangle$ will be affected. 

We can thus conclude that if $\delta V$ is gapped within $\mathcal H_{\text{Laughlin}}$, then $\Lambda^{ab}_{\text{Laughlin}}|\psi_0\rangle$ will also be gapped and there are two gravitons from the perspective of the spectral function, as is the case for the incompressible phase at $\nu=2/7$. If $\delta V$ is compressible, as is the case for the composite Fermi liquid state at $\nu=1/4$, $\Lambda^{ab}_{\text{Laughlin}}|\psi_0\rangle$ is also likely gapless or at very small energy. Only $\Lambda^{ab}_{LLL}|\psi_0\rangle$ will be gapped and experimentally measurable, at the same energy as other more familiar FQH phases such as the Laughlin state at $\nu=1/3$ or the Jain state at $\nu=2/5$.
 
\textit{Gravitons as edge excitations--} We have also explained that for any quantum Hall fluids, gravitons of both chiralities are present and are related by the inversion symmetry on the spherical geometry. This geometry is related to the disk geometry by a conformal transformation, and the disk geometry with an edge is also the most realistic geometry for the experimental measurement. Gravitons of opposite chiralities are also present on the disk geometry in principle, and we have shown that one chiral graviton has density modulation in the bulk (i.e. at the origin of the disk) and is thus gapped. For the Laughlin $\nu=1/3$ state this is the graviton given by $\hat B_{0,2}$, while for the anti-Laughlin state at $\nu=2/3$ it is given by $\hat B_{2,0}$. The graviton of the opposite chirality, on the other hand, does not lead to density modulation in the bulk, so the bulk is physically identical to the ground state. It is however physically equivalent to the quasihole state describing the density modulation at the edge of the quantum Hall fluid. For the sample size not too large compared to the area interacting with the incoming circularly polarized light, such neutral excitations in principle can also be detected at low energies, with a chirality that is opposite to that of the bulk graviton (see Fig.(\ref{diskraman})). The energy of such edge gravitons can be strongly dependent on the details of the confinement potential at the boundary of the quantum Hall fluid.

\section{Connection to other popular theories}\label{sec_connection}

In this section we first briefly comment on the description of the geometric modes in the fractional quantum Hall fluids, in particular the gravitons, from the perspective of the effective field theory and the composite fermion theory. For the last part, we will discuss in details another microscopic approach of constructing the chiral graviton modes using multi-body operators, in contrast to the one-body operator presented throughout this work. The description of the graviton modes of the strongly correlated quantum fluids with different languages and theoretical frameworks leads to important insights and intuition to the nature of such quantum fluids, as well as fundamental connections to other interesting physical system. It is however important note that we have well-defined microscopic construction of the geometric modes with transparent physical properties (e.g quantum numbers, sum rules, multiplicity and chirality of the geometric modes and even their energy scales), which should serve as the ``ground truth" for the phenomenological effective field theory and the CF description.

 \textit{Multiple gravitons from the effective field theory--} It was first proposed from the effective field theory description that for the integer quantum Hall states of composite fermions, additional neutral mode in the long wavelength may be needed so as not to violate the Haldane bound\cite{Nguyen2021}. Here the additional neutral mode, dubbed the Haldane mode, is put in by hand as additional degrees of freedom to satisfy the Haldane bound. The coupling of the quantum Hall fluids inelastically to photons is also derived in details from the effective field theory perspective \cite{Probing2021Nguyen}. In this context, the gravitons coupling to the photons are created by the stress tensor, which is the also a one-body operator effectively describing the $\rho(r)[\vec A(r)]^2$ in the microscopic theory. Here $\rho(r)$ is the unprojected bare density involving mixing of different LLs or even different bands in realistic materials, and $\vec A(r)$ is the external vector potential. 
 
It is not easy to understand from the effective field theory alone why the additional Haldane mode is needed, if and how many such modes are needed for a particular topological phase. All these require microscopic inputs and can be readily obtained from the formulation in the previous sections, especially in the language of the conformal Hilbert spaces\cite{wang2023geometric}. It is useful to understand why the Haldane bound can be violated in the effective field theory. It is implicitly assumed in the effective theory that the underlying degrees of freedom (or the elementary particles) are the CFs. The structure factor in the effective theory is thus implicitly computed from the basis of the CFs. In contrast, the topological shift of the FQH topological phase is computed from the electron basis within a single Landau level, and put into the effective theory by hand. Thus the former is the physical property within the CHS in which the CFs are the ``elementary fermions", while the latter is the physical property of the electrons. 

For example for CFs with an attachment of $2n$ fluxes per electron, they are the ``elementary fermions" within the CHS of the Laughlin $\nu=1/(2n+1)$, spanned by its ground states and quasiholes. This is because any electronic many-body wavefunctions for the quantum fluids of such CFs can be written as
\begin{eqnarray}\label{cfwf}
\psi\sim\prod_{i<j}\left(z_i-z_j\right)^{2n}\psi_{\text{CF}}
\end{eqnarray}
where $\psi_{\text{CF}}$ is a many-body wavefunction within the LLL so that $\psi$ is guaranteed to be of zero energy with respect to the pseudopotential interaction $\hat V_1+\hat V_3+\cdots\hat V_{2n-1}$. On the other hand, the topological shift is the physical quantity of the electrons in the CHS of a single LL. This inconsistency leads to the possibility of the violation of the Haldane bound and thus the necessity of the additional neutral modes.

Even with the inconsistency of the Hilbert spaces, it is not always the case that additional neutral modes are needed. This aspect of the physics cannot be captured by the effective theory alone, and depends on the algebraic structure of the CHS's. The number of additional neutral modes needed, in the language of the effective theory, also depends on the nature of the Hamiltonian, as we have described in details in Sec.~\ref{sec_dynamics} and more generally in Sec.~\ref{sec_chs}.

 \textit{Excitons in the CF theory--} As a very successful phenomenological theory, the CF theory and the related parton construction captures many essential features of the FQH fluids from the experimental measurements. More importantly, the theory allows for a very scalable numerical scheme to compute both universal topological properties and non-universal dynamical properties quite accurately for Coulomb interaction within the LLL. The fundamental assumption of the CF theory is that strongly interacting electrons in a partially filled band can be mapped to a weakly interacting CFs in a fully filled CF band; we can thus understand the FQH of electrons as the IQH of the CFs. The parton construction follows the same spirit, by partitioning a single CF into smaller fictitious components that form their own IQH, with the aim of capturing more FQH phases of electrons that cannot be understood as non-interacting CFs\cite{PhysRevB.40.8079}.

Naturally, the neutral excitations of the FQH fluids can be mapped to neutral excitations of the IQH of the CFs, which are excitons of CFs when one or more CFs are excited from the fully filled CF bands to the empty bands\cite{dev1992band,kamilla1996composite}. Interestingly even for the simplest exciton states in the CF picture, the mapping from the FQH of electrons to the IQH of the CFs is no longer exact: the $L=1$ single exciton neutral excitation of the FQH state does not exist due to translational invariance of the ground state. On the other hand the $L=1$ single CF exciton from the IQH picture is perfectly well defined, though the state vanishes after the prescribed LLL projection from numerical computations. This is one example where only the numerical evidence (thus not a rigorous statement) is present from the CF perspective why the $L=2$ graviton excitation is the long wavelength limit of the exciton spectrum. In contrast this is obvious from the symmetry argument in the microscopic picture.

While the CF and parton picture has the advantage of accurately capturing the dynamics of neutral excitations at both small and large wave-vectors with the same scheme (CF or parton excitons), one should note the neutral excitations at small wave-vectors are special: they emerge as quantized geometric fluctuations of the ground state thus qualitatively different from the dipole like finite wave-vector neutral excitations both in terms of symmetry and the relevant energy scales as discussed in Sec.~\ref{sec_intro}. Such geometric aspects of the FQH fluids are not explicitly captured by the CF or parton theory, but curiously it has been numerically established\cite{balram2024splitting,dora2024static,PhysRevLett.132.236503,PhysRevB.87.245125,PhysRevLett.107.086806,kamilla1996excitons, PhysRevLett.130.176501, PhysRevLett.108.256807} that in the long wavelength limit, the CF exciton and the GMP mode (thus the gravitons) have very large overlap for the primary Jain series, and an overlap of unity for the Read-Rezayi series including the Laughlin and Moore Read states. Nevertheless it may not be always easy to understand the the geometric modes within the CF and parton theory, and better understanding of the fundamental connections between the two perspectives are needed. 

For example with Coulomb interaction, the only graviton of the Laughlin state at $\nu=1/3$, and the high energy graviton of the FQH state at $\nu=2/7, 2/9$ and the CFL state at $\nu=1/4$ are physically the same type of graviton with the same chirality, very similar lifetime and variational energy (termed as the LL graviton in Sec.~\ref{sec_dynamics}). At $\nu=1/3$, this graviton can either be interpreted as an exciton of the composite fermion to the second CF level, or an exciton from the IQH of partons carrying a charge of $e/3$; at $\nu=2/7$ the same graviton is interpreted as an exciton from the IQH of partons carrying a charge of $2e/7$; at $\nu=2/9$ it is interpreted as an exciton from the IQH of partons carrying a charge of $2e/9$, while at $\nu=1/4$ it is interpreted as an exciton from the IQH of partons carrying a charge of $e/4$. From the clustering of the constructed first quantized wavefunctions, one can infer the energy of this graviton state will be punished by the $\hat V_1$ part of the interaction; however having many different interpretations of the same physical object can have both advantages and disadvantages. One needs to be a bit careful not to over-interpret the differences from these CF and parton constructions. From the geometric point of view the interpretation is straightforward: this graviton emerges from the quantum fluctuation of the guiding center metric outside of the Laughlin conformal Hilbert space ($\mathcal H_{\text{Laughlin}}$ in Sec.~\ref{sec_dynamics}); its energy thus comes dominantly from the $\hat V_1$ part of the Hamiltonian. Both statements are universal for all the FQH phases mentioned here. 

One should also note that not all trial wavefunctions of the exciton excitations from the parton theory are physical after the LLL projection and this requires testing from many-body numerical computations. Thus from the microscopic point of view, the CHS approach can again serve as the ground truth for the geometric aspects of the CF and parton trial wavefunctions, for example to understand if certain CF or parton constructions are physical graviton excitations for the quantum Hall fluid. On the other hand, as we can see from the recent work in anyon fractionalization, the geometric modes can serve as force mediating particles between charged excitations in quasihole fluids\cite{trung2023spin}. The reinterpretation of these modes as excitons from the CF picture can potentially lead to better understandings of various different roles such geometric modes can play in the dynamics of anyonic excitations in quantum Hall fluids.

 \textit{Multi-body chiral graviton operators--} While the one-body guiding center density operator (and its order by order expansion) gives a clear geometric interpretation for the excitations it induces, another insightful approach\cite{Liou2019,yang2016acoustic} is to look at the perturbation of the interaction Hamiltonian from a metric deformation. Without loss of generality we can take the metric deformation from $\eta=\left(1,0,0,1\right)$ to $\left(a,0,0,1/a\right)$ and assume the deformation is small: $a=1+\delta$. The effective two-body interaction given by $V\left(|\bm q|\right)\to V\left(|\bm q|\right)+\delta\left(q_x^2-q_y^2\right)\frac{V'\left(|\bm q|\right)}{2|\bm q|}$ up to the leading order in $\delta$. The perturbed interaction Hamiltonian is thus given by:
 \begin{eqnarray}\label{twobodyham}
 &&\hat H_{\text{int}}=\hat H_{0,\text{int}}+\delta\hat O^{\left(2\right)}\\
 &&\hat O^{\left(2\right)}=\int d^2q\left(q_x^2-q_y^2\right)\frac{V'\left(|\bm q|\right)}{2|\bm q|}\bar\rho_{\bm q}\bar\rho_{-\bm q}
 \end{eqnarray}
 
 Thus from an experimental point of view, if we are able to do a global deformation of the metric (e.g. by the acoustic wave), the excitation is given by the second term of Eq.(\ref{twobodyham}), from which we can compute the lifetime of the excitation from the spectral function. It is helpful to understand the second term from the complete basis of the generalized pseudopotentials\cite{PhysRevLett.118.146403} as follows:
 \begin{eqnarray}
 &&\hat O^{\left(2\right)}=\hat O^{\left(2\right)}_++\hat O^{\left(2\right)}_-, \hat O^{\left(2\right)}_-=\left(\hat O^{\left(2\right)}_+\right)^\dagger\\
 &&\hat O^{\left(2\right)}_+=\sum_M\sum_{m,n}c_{m,n}|m+n,M\rangle\langle m,M|\label{twobodydisk}\\
 &&c_{m,n}=\int d^2q\left(q_x^2-q_y^2\right)\frac{V'\left(|\bm q|\right)}{2|\bm q|}V_{m,n}\left(q_x,q_y\right)\\
 &&V_{m,n}\left(q_x,q_y\right)=\lambda_n \mathcal{N}_{mn} L_m^n\left(|q|^2\right)e^{-\frac{1}{2}|q|^2}\textbf{q}^n
 \end{eqnarray}
where $|m,M\rangle$ is a two-particle state with the relative angular momentum $m$ and the center of mass angular momentum $M$; the normalization factors are $\mathcal{N}_{mn}\equiv \sqrt{2^{n-2}m!/(\pi\left(m+n\right)!)}$, and $\lambda_n=1/\sqrt{2}$ for $n=0$ or $\lambda_n=1$ for $n\neq 0$. In particular Eq.(\ref{twobodydisk}) naturally applies for systems with rotational invariance, while for the torus we can use $\hat O^{\left(2\right)}_+=\sum_{m,n}c_{m,n}\int d^2q\lambda_n \mathcal{N}_{mn} L_m^n\left(|q|^2\right)e^{-\frac{1}{2}|q|^2}\textbf{q}^n\bar\rho_{\bm q}\bar\rho_{-\bm q}$. In realistic experiments, all coefficients of $c_{m,n}$ are generally non-zero and depend on the details of the realistic interaction when computing the linear response to the metric deformation. Theoretically, the chiral graviton operators are defined from a simple choice of a number of non-zero $c_{m,n}$ based on the model interaction Hamiltonian. If the model interaction Hamiltonian is three- or few-body interaction, the chiral graviton operators are also in the form of three- or few-body interactions and thus go beyond as a two-body operator.

Let us first discuss the similarities between the single-body graviton operators presented in this work, and the multi-body graviton operators previously proposed. Both are clearly geometry in nature. For the Laughlin model states at filling factor $\nu=1/\left(2p+1\right)$, the chiral graviton mode is given by $\hat B_{0,2}|\psi_0\rangle$ with the one-body operator. The corresponding two-body operator is given by $\hat O^{\left(2\right)}_-$ with only $c_{p,2}$ as the non-zero coefficient. It turns out with this operator, $\hat O^{\left(2\right)}_-|\psi_0\rangle=\hat B_{0,2}|\psi_0\rangle$ exactly so they are completely equivalent. Moreover, the corresponding $\hat O^{\left(2\right)}_+$ completely annihilates $|\psi_0\rangle$, showing the other chiral graviton does not exist. Analogously on the disk geometry $\hat B_{2,0}|\psi_0\rangle$ has exact zero energy with the model states, showing the other chiral graviton is gapless and solely due to the density modulation at the edge of the quantum Hall fluids.

With the one-body operator we can define the chiral graviton operators in different conformal Hilbert spaces, providing with a rigorous language for multiple gravitons. This can also be done with the multi-body operators, as long as such operator can be identified for a particular topological phase. Assuming $\hat O^{\left(2\right)}_{\pm}$ is identified, from the hierarchy of the CHS's given by Eq.(\ref{modelpham}) we can define $\hat O^{\left(2\right)}_{\pm, k}=\hat P_{k+1}\hat O^{\left(2\right)}_{\pm}\hat P_{k+1}-\hat P_{k}\hat O^{\left(2\right)}_{\pm}\hat P_{k}$. The determination of the number of gravitons based on the hierarchy of the CHS and the property of the ground state is completely analogous to the one-body graviton operators.

For the differences between the one-body and multi-body graviton operators, let us first discuss about the advantages of the multi-body operators. For the torus geometry, the multi-body operators are more convenient to construct the graviton states for finite systems, since for one-body operators, $\hat B_{m,n}$ themselves do not satisfy the periodic boundary conditions so we have to exponentiate them, and the long wavelength limit in principle can only be achieved in the thermodynamic limit. For the model wavefunctions of simple FQH states, the multi-body operators can tell us right away if one chiral graviton will vanish from the clustering of the ground state wavefunction\cite{bernevig2008generalized, PhysRevB.75.075318}. For one-body operators, in principle we need to either check the chiral graviton energy to see if they are edge excitations, or the density modulation of the graviton states on the sphere; both require numerical computations. On the other hand even for FQH phases in general, the chirality of the graviton modes can be rather easily determined from the property of the ground state without any numerical computations, given that the chirality is odd under particle-hole conjugate. This argument is independent of which graviton operator is used.

The advantages of the single particle graviton operators include their generality: one-body operators are universal for any FQH states, in contrast to multi-particle operators that depend on the (model) Hamiltonian, so they are different for different FQH states. The generalization of the one-body graviton operators to higher spin one-body geometric operators is natural and straightforward from a geometric and algebraic point of view (see Sec.~\ref{sec_lwl}), from $\hat B_{0,2}$ and $\hat B_{2,0}$ to $\hat B_{m,n}$. It is not entirely clear how this can be done for multi-particle operators, but presumably this can also be achieved from the higher order expansion in $\delta$ in Eq.(\ref{twobodyham}). From the numerical computation point of view (especially for exotic FQH phase), one-body operators are also much more resource efficient as compared to multi-body operators. 

From the experimental point of view, for Raman scattering with circularly polarized light, photons couple to the GMP modes (i.e. the density modulation, see Fig.(\ref{fig1})) non-linearly\cite{wurstbauer2013resonant,pinczuk1998light,pinczuk1993observation,platzman1996resonant,kang2001observation,kukushkin2009dispersion,du2019observation,wurstbauer2015gapped,pinczuk1994inelastic,liang2024evidence}, and thus the one-body graviton operators are the operators of choice to predict both qualitatively and quantitatively experimental measurements. This is particularly true when photons do impart a small transfer of momentum; thus when coupling to the GMP mode in the long wavelength limit, the symmetry of the ground state is important. The graviton modes may always exist in the system with long enough lifetime, but the GMP modes are saturated with dipole like excitations if the ground state breaks translational symmetry (see Sec.~\ref{sec_lwl}), making Raman scattering not suitable for detecting the gravitons. On the other hand this makes Raman scattering a useful tool for detecting symmetry-breaking quantum Hall fluids. The multi-body operators are specifically for perturbations that leads to a uniform change of the metric of the Hall manifold. This can come from for example the quench of the quantum Hall fluids by tilting of the magnetic field of a 2D electron gas with a finite thickness\cite{liang2024evidence}, or with acoustic waves perpendicular to the Hall manifold\cite{yang2016acoustic} (with no momentum transfer). It is not clear, however, how chiral gravitons can be excited in these approaches, in contrast to Raman scattering. 

It is also important to note that as long as the ground state is translationally invariant, the long wavelength limit of the GMP mode always allow the coupling of the chiral gravitons to the photons, where such gravitons correspond to the single particle graviton operators. For linear response to the metric deformation, however, it will only excite the graviton if the interaction between electrons are the artificial model interactions. Using the Laughlin state at $\nu=1/3$ as an example, a metric deformation only gives the chiral graviton if the interaction is the $\hat V_1$ pseudopotential, since the linear response gives $\hat O^{\left(2\right)}_{\pm}$ with only $c_{1,2}$ as the non-zero coefficient. With Coulomb interaction, even though the ground state hardly changes, the metric deformation leads to $O^{\left(2\right)}_{\pm}$ with all $c_{m,2}$ non-zero for $m$ odd. Thus in addition to the chiral gravitons, many other neutral excitations are induced by the metric deformation, and in particular $\hat O^{\left(2\right)}_+$ no longer annihilates the ground state, and instead give resonant peaks just like $\hat O^{\left(2\right)}_-$ with the Coulomb interaction. 

\begin{acknowledgments}
I thank Ajit Balram, Jainendra Jain, Dung Xuan Nguyen, Lingjie Du and Kun Yang for the very constructive discussions, and Yuzhu Wang for the illustrations in this work. This work is supported by the NTU grant for the National Research Foundation, Singapore under the NRF fellowship award (NRF-NRFF12-2020-005), and Singapore Ministry of Education (MOE) Academic Research Fund Tier 3 Grant (No. MOE-MOET32023-0003) “Quantum Geometric Advantage.”
\end{acknowledgments}

\bibliography{ref}

\end{document}